\documentclass[final]{aa}
\usepackage[varg]{txfonts} 
\usepackage{pgfplots}
\usepgfplotslibrary{external}
\usepackage{graphicx}
\usepackage{grffile}
\usepackage{units}

\usepackage[utf8]{inputenc}

\usepackage{amsmath}
\usepackage{amssymb}

\usepackage{caption}
\usepackage{subcaption}
\usepackage{units}

\usepackage{natbib}
\bibpunct{(}{)}{,}{a}{}{,} 

\usepackage{multirow}
\usepackage{tabularx}

\usepackage{import}

\newcommand{\LEt}[1]{}

\DeclareGraphicsExtensions{.pdf,.png,.jpg}

\renewcommand{\phi}{\varphi}

\newcommand{\dd}{\,\mathrm{d}}

\newcommand{\sfrac}[2]{#1/#2}

\newcommand{\figpgf}[5][1.0]{\begin{figure}
\centering

\renewcommand\pgfimage[2][NOSTD]{\includegraphics[##1]{#2##2}}\resizebox{#1\linewidth}{!}{\input{#2#3.pgf}}
\caption{#5}
\label{#4}
\end{figure}
}

\newcommand{\figtwovpgf}[7]{\begin{figure}
\centering

\renewcommand\pgfimage[2][NOSTD]{\includegraphics[##1]{#4##4}}\resizebox{#1\linewidth}{!}{\input{#4#4.pgf}}

\renewcommand\pgfimage[2][NOSTD]{\includegraphics[##5]{#5##5}}\resizebox{#5\linewidth}{!}{\input{#5#5.pgf}}
\caption{#7}
\label{#6}
\end{figure}
}

\newenvironment{stabular}[2][1.2]
{\def\arraystretch{#1}\tabular{#2}}
{\endtabular}

\let\originaleqref\eqref
\renewcommand{\eqref}{Eq.~\originaleqref}

\usepackage{changes}

\begin{document}

\author{M.~Jung\inst{\ref{HS},\ref{CAU}} 
\and T.~F.~Illenseer\inst{\ref{CAU}} 
\and W.~J.~Duschl\inst{\ref{CAU},\ref{ARIZONA}}}

\institute{
Institute for Theoretical Physics and Astrophysics, Kiel Astrophysics, Christian-Albrechts-University Kiel, Leibnizstraße 15, D-24118 Kiel, Germany\label{CAU}
\and
Current affiliation: Hamburger Sternwarte, Universität Hamburg, Gojensbergweg 112, D-21029 Hamburg, Germany, \\\email{manuel.jung@hs.uni-hamburg.de}\label{HS}
\and
Steward Observatory, The University of Arizona, Tucson, AZ 85721, United States\label{ARIZONA}
}

\newcommand{\refree}[1]{{#1}}
\newcommand{\refreec}[1]{{#1}}
 
\title{Multi-scale simulations of black hole accretion in barred galaxies: Self-gravitating disk models.}


\abstract{
Due to the non-axisymmetric potential of the central bar,  in addition to their characteristic arms and bar, barred spiral galaxies
form a variety of structures within the thin gas disk, such as nuclear rings, inner spirals, and dust lanes. These structures in the inner kiloparsec are extremely important in order to explain and understand the rate of black hole feeding. The aim of this work is to investigate the influence of stellar bars in spiral galaxies on the thin self-gravitating gas disk. We focus on the accretion of gas onto the central supermassive black hole and its time-dependent evolution. We conducted multi-scale simulations simultaneously resolving the galactic disk and the accretion disk around the central black hole. In all the simulations we varied the initial gas disk mass. As an additional parameter we chose either the gas temperature for isothermal simulations or the cooling timescale for non-isothermal simulations. Accretion was either driven by a gravitationally unstable or clumpy accretion disk or by energy dissipation in strong shocks. Most of the simulations show a strong dependence of the accretion rate at the outer boundary of the central accretion disk ($r<\unit[300]{pc}$) on the gas flow at kiloparsec scales. The final black hole masses reach up to
 $\sim\unit[10^9]{M_\odot}$ after $\unit[1.6]{Gyr}$. Our models show the expected influence of the Eddington limit and a decline in growth rate at the corresponding sub-Eddington limit.
}

\maketitle
\section{Introduction}
Recent very high-resolution observations (${\sim}\unit[100]{pc}$) have given the first insights into the inner structures of barred spiral galaxies and reveal the circumnuclear disk, nuclear spirals, and rings, which are often subject to high star formation rates \citep{2015ApJ...799...11X,2015ApJ...806L..34F,salak_gas_2016}. These efforts gradually increase our knowledge about this kind of galaxies on which detailed models can be built. Despite these improvements we still lack a    consistent picture of the angular momentum transport mechanisms and the feeding of the central supermassive black hole. In case of active galactic nuclei (AGN), measurements of the nuclear luminosity and the bulge velocity dispersion allowed us  to make a connection between the black hole and the bulge early on \citep{2000ApJ...539L...9F,2000ApJ...539L..13G,2011Natur.480..215M}. While an explanation for this connection is still highly debated, the importance of multiple different heating and cooling processes such as star formation, supernova explosions, and AGN feedback cannot  be neglected. An excellent review highlighting the various aspects of black hole growth can be found in \citep{2012NewAR..56...93A}.\\
In the last 20 years there have been great efforts to include different effects of the energy balance in galaxies. In regions with high star formation rates the starlight can heat the gas \citep{2006MNRAS.373.1074S,2010ApJ...717..121C}, and the stars rapidly explode as supernovae and inject momentum and energy into the interstellar medium \citep{1997MNRAS.284..235Y,2008MNRAS.389.1137S,2010MNRAS.405.1491M}. Also, there have been efforts to include the observed multi-phase medium into the models \citep{2003MNRAS.339..289S,2012MNRAS.421.1838B}. All of these simulations are based on the smoothed particle hydrodynamics (SPH) method \citep{gingold_smoothed_1977,lucy_numerical_1977}. \refree{Resolving a self-gravitating accretion disk in the central nuclear region ($<\unit[300]{pc}$) with the SPH method is difficult. Although this region can have a high gas density, it contributes only  $2\%$ to the overall mass of the gaseous disk. Thus, the number of SPH particles in this region will only be about $2\%$ of all particles. Achieving a high number of neighbor particles and high resolution in this region would require spending most particles on out-of-focus regions.}\\
There have been attempts to cover multi-scale gas behavior by stopping and restarting the simulation at a higher resolution \citep{2010MNRAS.407.1529H},  but ultimately high resolution in the central region can only be achieved with grid-based methods. Because of the high velocities near the black hole and the inclusion of self-gravity, this  is very
computationally expensive even in 2D.\\
Several 2D simulations of barred galaxies have been carried out before. \refree{Early simulations were focused on the large-scale  structures (e.g., density waves) and their dependence on pattern speeds \citep{1979A&A....78..133B,1988MNRAS.231P..25S,1996A&A...313...65L}.}
 \cite{1992MNRAS.259..345A} discussed the dependence of the shape and position of dust lanes on general galaxy parameters \LEt{  the dependence of x ON y; x depends ON y } (e.g., the bar parameter) and found substantial inflow in models with strong shocks. Later \cite{athanassoula_gas_2000} confirmed \refree{these} findings and briefly discussed the role of a secondary bar inside of the main one. In addition,  \cite{maciejewski_gas_2002} examined the role of secondary bars and found that they cannot generate shocks in the gas flow, which would be visible in observations and are unlikely to increase mass inflow into the galactic nucleus. The curvature of the dust lanes is determined by \cite{comeron_curvature_2009} and they deduce that it is predominantly a result of the parameters of the bar. \cite{sormani_gas_2015a} investigate the gas streamlines for different orbit families in detail and find for sufficiently high sound speeds and resolution that the flow downstream of the shocks becomes unsteady.\\
The main focus of most publications has been on nuclear structures like nuclear rings and spirals. The generation and position of nuclear rings was analyzed by  \cite{1995ApJ...449..508P} and they found a connection to the inner Lindblad resonance position. \cite{regan_formation_2003} disagreed and provided a model in which they are not related to inner Lindblad resonances as was previously thought. \cite{namekata_mass_2009} discuss the impact of nested bars on the gas inflow and find that a nuclear self-gravitationally unstable gas disk is generated for some simulation setups. \cite{2012ApJ...747...60K} survey the dependencies of the structure of nuclear rings and spirals to the sound speed and the mass of a central black hole. \cite{sormani_gas_2015c} vary the quadrupole of the bar potential and compare it to the longitude--velocity diagram of the Milky Way to constrain bar parameters. \cite{2015ApJ...806..150L} distinguish between round and elongated nuclear rings and identify global parameters, which correlate with the position of round rings.\\
Several publications are dedicated to the effects of bar strengths. \cite{regan_bar-driven_2004} find that weak bars have almost no effect on the radial distribution of the gas, but strong bars can drive gas down to the nuclear ring, but not any further. \cite{kim_gaseous_2012} investigate the complicated dependencies of nuclear structures on different bar strengths. The large-scale formation of the spiral arms emerging from the ends of galactic bars are examined by \cite{sormani_gas_2015b}. They find that they can be understood as kinematic density waves.\\
Magnetic fields can enhance the mass influx rate driven by shocked gas in dust lanes. Eventually a magnetic dynamo outside of the corotation resonance radius can turn the outer disk into a highly chaotic state  \citep{2012ApJ...751..124K,kim_magnetohydrodynamic_2013}. Star formation in nuclear rings, which is solely fueled by gas inflow as the result of a bar potential (which was  examined by \citep{2013ApJ...769..100S,kim_formation_2014}) occurs mainly at the connection point between nuclear ring and dust lanes. The star formation rate correlates with the mass inflow rate to the ring, but not with the mass of the ring.\\
In this work we focus on the black hole accretion flow and extend the simulation domain to the subparsec scale. We also account for self-gravity of the gas.\\
In the second part of this paper we eliminate the restriction on isothermal models. We account for the contribution of shock and compressional heating and a simple one-parameter cooling depending only on the orbital timescale are considered. The selection of the cooling parameter in this connection is based on the critical value of $b_\mathrm{cool}$ \citep{2001ApJ...553..174G}, which is known to produce marginal stable accretion disk flows in these systems. This is also justified by the observations of \cite{2015ApJ...806L..34F}, who give an estimate for the Toomre parameter \added{\citep{1964ApJ...139.1217T}} of NGC 7469\deleted{ (???)}, from which follows that the inner region should be gravitationally unstable. These models are also the first steps  toward more detailed models of the heating and cooling effects,  among \LEt{ You show a preference for US spelling conventions, and I have corrected accordingly throughout.} others, include stellar radiation feedback, star formation feedback, and supernova explosions, which is the scope of future work.\\
In the first paper \citep{Jung2017a} (hereafter Paper I) we established the numerical foundation to execute isothermal multi-scale simulations of black hole accretion in barred galaxies. To do this, we introduced a method for conserving \refree{inertial} angular momentum in a rotating reference frame, a spectral self-gravity solver and discussed the importance of the gravitational energy transport in the energy equation. \refree{These extensions are crucial in order to accurately simulate the accretion onto the central black hole, which is significantly determined by the angular momentum transport processes within the disk\refreec{, and  is the subject of this paper}.}\\
The paper is structured as following. In section 2 we present the model of a self-gravitating gas disk subject to multiple gravitational potentials. In section 3 we present our isothermal simulation results and classify them in a strong and a weak accretion mode. In section 4 we outline the results from the simulations with cooling, and classify them in a hot and cold accretion mode. In section 5 we present our conclusions.

\section{Model}\label{sec:numsetup}
\subsection{Gravitational potentials}
The most important part of all following galaxy simulations is the implementation of the gravitational potential. 
The set of potentials applied in this work is basically that given in   \cite{2012ApJ...747...60K}, although we modified their implementation by adding a Newtonian point mass potential for the black hole and relaxing the razor thin disk approximation for the stellar disk potential. The parameters in the gravitational potentials are chosen to be similar to \cite{1995ApJ...449..508P} and \cite{2012ApJ...747...60K}, thus a typical flat galaxy rotation profile is recovered.
\paragraph{\textbf{Black hole}}
The black hole is modeled by a Newtonian point mass potential:
\begin{equation}
\Phi_\text{BH} = -\frac{G M_\text{BH}}{R}.
\end{equation}
The inner boundary of the simulation domain is far enough away to neglect any relativistic effects. 
The initial mass of the black hole $M_\mathrm{BH}=\unit[10^6]{M_\odot}$ is allowed to grow during the simulation.
The mass flux across the inner boundary is calculated at each time step, which sometimes yields super-Eddington accretion rates. In that case we limit the actual black hole accretion rate $\dot{M}_\textsf{BH}$ to  at most the Eddington rate $\dot{M}_\textsf{Edd}$ \citep{2011ApJ...728...98D}. At each time step $\dot{M}_\textsf{acc}$ is calculated and the Eddington limited result of this is added to the black hole mass:
\begin{equation}
\dot{M}_\textsf{BH} = \begin{cases}
\dot{M}_\textsf{acc} & \text{if } \dot{M}_\textsf{acc} \leq \dot{M}_\textsf{Edd}\\
\dot{M}_\textsf{Edd} & \text{else}.
\end{cases}
\end{equation}
If the accretion rate is greater than the Eddington limit, the excess mass is removed from the system. We call this Eddington outflow:
\begin{equation}
\dot{M}_\textsf{outflow} = \dot{M}_\textsf{acc}-\dot{M}_\textsf{BH}.
\end{equation}
While feedback by the Eddington outflow is not considered in these simulations, 
radiation pressure onto the gas disk is negligible since the disk is geometrically thin. A feedback by the black hole spin is part of the black hole accretion efficiency factor $\eta$ \citep{2011ApJ...728...98D}.
\paragraph{\textbf{Stellar disk}}
The stellar disk is approximated by a Kuzmin--Toomre model \citep{kuzmin1956model,1963ApJ...138..385T,2008gady.book.....B} for the stellar density, which is given by
\begin{align}
\rho_\mathrm{disk}\left(r,z=0\right) = &\frac{R_1^2 M_\mathrm{disk}}{4\pi}\nonumber\\
&\cdot\frac{R_0 r^2 + \left(R_0+3\cdot R_1\right)\cdot\left(R_0+R_1\right)^2}{R_1^3 \left(r^2+\left(R_0 + R_1\right)^2\right)^{\frac{5}{2}}}.
\end{align}
The corresponding midplane potential is
\begin{equation}
\Phi_\text{disk}\left(r,z=0\right) = -\frac{G\cdot M_\mathrm{disk}}{\sqrt{r^2+(R_0+R_1)^2}}
\end{equation}
with the constants $R_0=\unit[14.1]{kpc}$ and $R_1=\unit[14.1]{pc}$. The total disk mass amounts to
\begin{equation}
M_\mathrm{disk} = \frac{v_0^2 R_0}{G} = \unit[2.2\cdot 10^{11} ]{M_\odot}
\end{equation}
with $v_0 = \unit[260]{km/s}$.
\paragraph{\textbf{Bulge}}
The bulge is modeled by a modified spherically symmetric Hubble profile \citep{binney_galactic_1987} of the density
\begin{equation}
\rho_\mathrm{bul}\left(r\right) = \rho_{\mathrm{bul},0}\left(1+\frac{r^2}{R_\text{b}^2}\right)^{-3/2}
\end{equation}
resulting in the potential
\begin{equation}
\Phi_\text{bul}\left(r\right) = -\frac{4\pi G\rho_\text{bul} R_\text{b}^3}{R}\ln\left(\frac{r}{R_\text{b}}+\sqrt{1+\frac{R^2}{R_\text{b}^2}}\right).
\end{equation}
The density is $\rho_{\mathrm{bul},0}=\unit[2.4\cdot 10^{10}]{\frac{M_\odot}{kpc^3}}$ and the characteristic extension $R_b=\unit[0.33]{kpc}$.
The mass inside of a $\unit[6]{kpc}$ radius of the bulge amounts to $M_\mathrm{bul}=\unit[2.8\cdot 10^{10}]{M_\odot}$.
\paragraph{\textbf{Bar}}
The bar potential is modeled using a Ferrers potential \citep{ferrers1887,1969efe..book.....C,1984A&A...134..373P} for the stellar density
\begin{equation}
\rho_\mathrm{bar} = \begin{cases}\rho_{\mathrm{bar},0}\left(1-g^2\right)^n &\mbox{if } g < 1,\\ 0 &\mbox{\replaced{else}{sonst}.\LEt{ otherwise? else? }}\end{cases}
\end{equation}
with $g=\sfrac{x^2}{a^2}+\sfrac{y^2}{b^2}+\sfrac{z^2}{c^2}$;
\replaced{$a$ the major and $b$, $c$ the minor semi-axes}{$a$ and $b$, $c$ are the major and minor semi-axis}, respectively.\LEt{ a, b, and c?  three variables and  two semi-axes? (so can't be respectively). or do you mean a is the major semi-axis and b and c are the two minor  semi-axes? see also note 26   }
The Ferrers potential can be used for a general ellipsoid, if specifying the third semi-axis length $c\neq b$.
We set in these simulations $c=0.99b$ since the Ferrers potential is only defined for $c < b$, but we want to achieve approximately a prolate spheroid.
The fourth parameter $n$ in the exponent controls the decline of the density. We set for all the simulations in this paper $n=1$, $a=\unit[5]{kpc}$, $b=\unit[2]{kpc,}$ and $\rho_\mathrm{bar}=\unit[0.45]{\frac{M_\odot}{pc^3}}$.
The total mass of the bar is $M_\mathrm{bar}=\unit[1.5\cdot 10^{10}]{M_\odot}$.
The midplane bar potential is
\begin{align}
\Phi_\text{bar}\left(x,y,0\right) = &-\frac{\pi G a b c \rho_\text{bar}}{2}\nonumber\\
&\cdot\left(W_{000}+x^2\left(x^2 W_{200} +2 y^2 W_{110}-2 W_{100}\right)\right.\nonumber\\
&\quad\left.+y^2\left(y^2 W_{020}-2 W_{010}\right)\right),
\end{align}
using the definition of the potential and its coefficients $W_{ijk}$ from \cite{1984A&A...134..373P}.
The  coefficients  are repeated in  Appendix~\ref{sec:barpotential}.
The bar potential is slowly switched on during the first rotation of the bar, so that a violent adoption of the gas surface density can be prevented.
To suppress radial perturbations entirely, we keep the total system mass constant. The initial bulge mass is the sum of the intended bulge and bar masses. During the switch-on phase mass is removed from the bulge at the same rate as it is added to the bar \citep{2012ApJ...747...60K}. Therefore, the rotation law as well as the balance of centrifugal force and radial gravitational forces remain unchanged.
The bar rotates with the angular velocity
\begin{equation}
\Omega_\mathrm{p} = \unit[33]{\frac{km}{s\cdot kpc}}.
\end{equation}
Since a rotating frame of reference is used with an angular velocity of $\Omega_\mathrm{p}$, the bar rests in all simulations along the x-axis.\\

All of these potentials model the gravitation of stars, dark matter, and of a black hole.
In addition to the discussed gravitational components, we allow for self-gravitation \citep{chan_spectral_2006,li_fast_2009}.
Hereafter we use the delta distribution as Green function to model a razor thin gas disk.

\subsection{Initial and boundary conditions}\label{sec:initialconditions}
In the simulations we consider the gas component as the vertically integrated surface density $\Sigma$.
The disk is discretized as a mesh in polar coordinates using a logarithmic radial scaling.
The radial extent spans $R\in[\unit[2]{pc},\unit[16]{kpc}]$. 
The mesh resolution is $N_r\times N_\phi = 384 \times 256$, which amounts to an aspect ratio of $\sim 1$ within each cell \refree{and to a cell size $\Delta x$ of $\unit[0.05]{pc}$, $\unit[0.25]{pc}$, and $\unit[25]{pc}$ at $\unit[2]{pc}$ (inner rim), $\unit[10]{pc,}$ and $\unit[1]{kpc}$, respectively.
Due to the logarithmic radial scaling, half of the radial cells resolve the inner accretion disk $r < \unit[300]{pc}$, which is the most important region for the accretion flow. We trade some overall resolution to cover for the extended cost and number of \replaced{shorter}{smaller} time steps,\LEt{ smaller number of time steps ? // Is CFL an abbreviation that needs to be introduced? All abbreviations/acronyms (and instrument/telescope names where appropriate) need to be introduced at first use, once in the Abstract and then again in the body of the paper; the abbreviation should then be used consistently. Please check throughout the paper.} because of the increased demands of the CFL criterion \citep{courant_uber_1928} in connection with the small cells at the inner boundary to ensure numerical stability, very long simulation run times and self-gravity. One simulation leverages about two weeks of real time on the provided NEC SX ACE vector super computer. This amounts to $\unit[8064]{CPU}$ hours for each simulation setup, which is roughly equivalent to $\unit[130000]{CPU}$ hours on Intel Xeon E5-2670 CPUs.}\\
The initial surface density follows a power law with an exponent $\kappa=-1$, which is limited near the inner boundary:
\begin{equation}
\Sigma\left(r\right) = \Sigma_0 \cdot \min\left(\left(\frac{r}{R_\kappa}\right)^\kappa, 1\right).
\end{equation}
The  cutoff radius is $R_\kappa=\unit[100]{pc}$ and the initial mass $\Sigma_0$ is a parameter of the model space. With this cutoff radius we achieve similar mass densities at $R=\unit[1]{kpc}$ as in \cite{2012ApJ...747...60K}.
A power law with negative exponent is necessary for the surface density, because otherwise a local minimum of the gravitational potential, generated by the self-gravitation, would lead to non-physical solutions.
\cite{2015MNRAS.450..691I} describe in detail which power laws are sensible choices for accretion disks.
Our choice for the power law and the initial density generate densities at radius $R_\kappa$, which are comparable to those \deleted{generated }in  \cite{2012ApJ...747...60K}.
To excite any potentially existing instabilities, we add a noise on the order of $10^{-3}$ to the surface density distribution.
Table~\ref{tab:agnmassen} summarizes the initial mass of all gravitational sources.\LEt{ Single-sentence paragraphs should be avoided. Can this sentence be included with the previous or following paragraph? Please check throughout.  }\\
\begin{table}\centering
        \caption{Summary of the initial masses of the different gravitational components and initial surface density parameters.}
        \begin{stabular}[1.5]{rcccc}\hline\hline
                density $\Sigma_0$ & {[$\unit{M_\odot/pc^2}$]} & $10$ & $30$ & $50$ \\%\hline
                gas disk     & [$\unit{M_\odot}$] & $1.0 \cdot 10^9$ & $3.0 \cdot 10^{9\phantom{0}}$ & $5.0 \cdot 10^9$ \\\hline
                stellar disk & [$\unit{M_\odot}$] & \multicolumn{3}{c}{$2.2\cdot 10^{11}$} \\%\hline
                bulge & [$\unit{M_\odot}$] & \multicolumn{3}{c}{$2.8\cdot 10^{10}$} \\%\hline
                bar & [$\unit{M_\odot}$] & \multicolumn{3}{c}{$1.5\cdot 10^{10}$} \\%\hline
                black hole & [$\unit{M_\odot}$] & \multicolumn{3}{c}{$1.0\cdot 10^{6\phantom{0}}$} \\
        \end{stabular}
        \label{tab:agnmassen}
\end{table}
At the beginning of the simulation the radial velocity field is zero and the azimuthal velocity is chosen so that the centrifugal force is balanced by all other forces, i.e., gravitational and pressure gradient forces.
Numerically, this is achieved by evaluation of the right-hand side of the system of differential equations, which consists of all source terms and the flux difference quotients. This is important because generally it will deviate from the analytic term by  means of the numerical flux calculation and linear reconstruction. A more detailed explanation can be found in Paper I.
The resulting rotation curve shows the typical flat slope at large radii (see \citealp{1985ApJ...295..305V}). The difference \replaced{to}{with} curves for simulations of different total gas mass and therefore stronger self-gravitation is negligible. The gas rotates in the mathematical positive direction, i.e.,  in the rotating reference frame within (beyond) the corotation radius counterclockwise (clockwise).\\
\refreec{At the outer boundary we use a no-gradients condition. At the inner boundary we set an outflow condition\refree{ and extra\-polate the azimuthal velocity with a Keplerian profile to prohibit gas inflow into the computational domain}. In general when simulating accretion disks one has to choose how to extrapolate the specific angular momentum in the boundary cells. While  a no-gradients boundary condition is often chosen as an approximation at the inner boundary, it can induce a torque into the inner disk that can alter the flow significantly. According to our experience a Keplerian extrapolation \replaced{introduces}{induces} no noteworthy torque \replaced{into}{in} the inner disk.\LEt{ introduces ... into   }
Strictly speaking an inner Kepler boundary condition is only a good approximation if the rotation profile in a few cells next to the inner boundary is also Keplerian during the entire simulation time. In particular, this requires an initially non-self-gravitating inner accretion disk, otherwise the surface density distribution would dictate a  non-Keplerian rotation profile. However,  without a density cutoff the mass of the disk $r<\unit[100]{pc}$ would approach an overall mass  on the order of the initial black hole mass ($\unit[10^6]{M_\odot}$) and has to be considered self-gravitating. At the same time $\Sigma_0$ is fixed since the surface density at $\unit[1]{kpc}$ should be comparable to the value of \cite{kim_gaseous_2012}. All the simulations in this paper indeed show a rotation profile in a few cells next to the inner boundary, which is close to Keplerian throughout the complete simulation time.}\\
One  rotation of the bar and the orbital periods at the inner and outer boundary last respectively
\begin{align}\label{eq:dynscales}
t_\mathrm{bar} &= \unit[186]{Myr},\nonumber\\
t_\mathrm{dyn,inner} &= \unit[0.3]{Myr},\\
t_\mathrm{dyn,outer} &= \unit[500]{Myr}.\nonumber
\end{align}
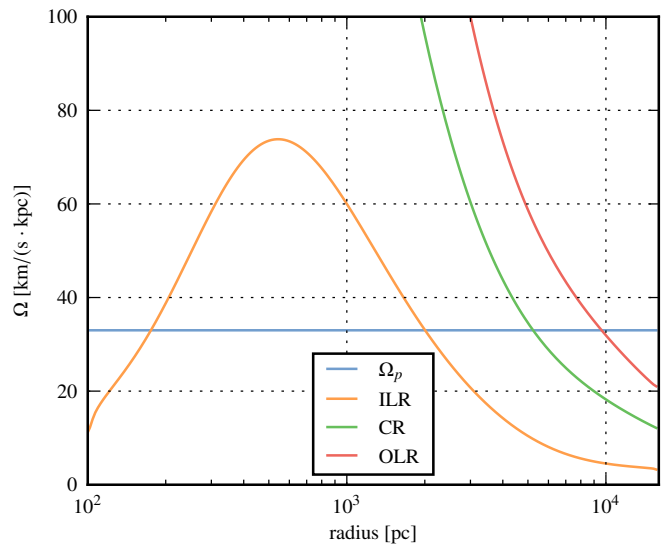
\figpgf{sims/figure/agndiskisoflat/1.E+1_4.5E-1_0.E-0_1.E+3/}{lindblad_0000}{lindblad}{ Frequency distribution curves  of the inner Lindblad (ILR), outer Lindblad (OLR), and  corotation (CR) frequency. The horizontal line is the bar frequency $\Omega_\mathrm{p}$. The intersections of these curves are resonance loci.}
The simulation stop time is after nine bar rotations
\begin{equation}
t_\mathrm{sim} = \unit[1674]{Myr}.
\end{equation}
\refree{Lindblad resonances are important for the dynamics of the system \citep{1964ApJ...140..646L,1966PNAS...55..229L,2008gady.book.....B}). 
Figure~\ref{lindblad} shows their dependencies on the initial resonant frequencies and the bar angular velocity for one of the simulations.}
At the beginning this structure is very similar in all the simulations because all the gravitational potentials remain the same except for the contribution from the self-gravitating disk.
The locations of the inner and outer Lindblad resonances, $r_\mathrm{ILR}$ and $r_\mathrm{OLR}$, respectively, as well as the corotation resonance $r_\mathrm{CR}$ are given by
\begin{align}
r_\mathrm{ILR} &= \unit[\left\{42,173,1970\right\}]{pc},\nonumber\\
r_\mathrm{CR} &= \unit[5.14]{kpc},\\
r_\mathrm{OLR} &= \unit[9.45]{kpc}.\nonumber
\end{align}
A constant speed of sound is defined for the isothermal simulations in the entire calculation domain, which is the second parameter of the model space.
If we assume an ideal, completely ionized gas consisting of $75\%$ hydrogen and $25\%$ helium, the equation of state is given by
\begin{equation}
p =\left(\gamma-1\right)\rho e
,\end{equation}
where the mean molecular weight can be specified as $\mu=\unit[6.02\cdot 10^{-4}]{kg/mol}$ \citep{kippenhahn_stellar_1990}.
The temperature is then a function of gas density and pressure: 
\begin{equation}\label{eq:trhop}
T = \frac{\mu}{R_\mathrm{G}} \frac{p}{\rho}.
\end{equation}
Here $R_\mathrm{G}$ identifies the universal gas constant. The isothermal equation of state
\begin{equation}
\frac{p}{\rho} = c_\mathrm{s}^2
\end{equation}
is used for the isothermal simulations. Therefore, in this case the temperature is a function of the isothermal sound velocity
\begin{equation}\label{eq:temp}
T = \frac{\mu}{R_\mathrm{G}} c_\mathrm{s}^2.
\end{equation}

\subsection{Non-isothermal extensions}\label{sec:numsetupext}
For the non-isothermal simulations, we use the parameterized cooling function  proposed by \cite{2001ApJ...553..174G}.
It is based on the coupling of the cooling timescale $\tau_\mathrm{cool}$ to the dynamical timescale $\tau_\mathrm{dyn}$ by the dimensionless parameter $b_\mathrm{cool}$
\begin{equation}
\tau_\mathrm{cool} = b_\mathrm{cool} \cdot \tau_\mathrm{dyn}.
\end{equation}
Using the internal energy $e=p/(\gamma-1)$ and the azimuthal angular velocity $\Omega$ the cooling source term $S_\mathrm{cool}$ becomes
\begin{equation}
S_\mathrm{cool} = -\frac{e\Omega}{b_\mathrm{cool}}.
\end{equation}

Irrespective of hydrodynamic heating processes (e.g., shock and compressional heating), we model no further heating mechanisms. In addition,  gas in spiral galaxies can be heated by star formation and the resultant stellar winds and supernovae \citep{1977ApJ...218..148M,1997MNRAS.284..235Y,2003MNRAS.339..289S}.
If we considered such feedback mechanisms, a more realistic cooling would be required (e.g., the method from \cite{1990ApJ...351..632H} based on gray cooling with Rosseland mean opacities \citep{1994ApJ...427..987B}).
A cooling of this nature operates on timescales on the order of $\unit[10^{-6\dots -4}]{\tau_\mathrm{dyn}}$ \citep{2010MNRAS.407.1529H}, which is considerably shorter than the typical hydrodynamical  time step $\unit[{[10-100]}]{yr}$.
The number of required time steps would rise  by a factor of $10-100$ when using a detailed model for the cooling.
As a consequence it would be too computationally expensive to follow the simulations during multiple bar rotations since the present parallel calculations already leverage the available computing time of two weeks of real time per simulation (see section~\ref{sec:initialconditions}).\\
The estimation of the heating timescale is \added{difficult} due to the numerous processes involved. The lifetimes of the mass-rich stars, which are \replaced{the most}{very} important for the energy input, is typically less than 
$\unit[10^7]{yr}$ \citep{1997MNRAS.284..235Y} before they release a huge amount of energy by means of a supernova explosion.
Simulations of the complete star formation cycle are so substantial and complex  that even the most comprehensive hydrodynamical models have to \replaced{simplify them}{be simplified} \citep{1997MNRAS.284..235Y,2003MNRAS.339..289S,2006MNRAS.373.1074S,2010ApJ...717..121C,2008MNRAS.389.1137S,2012MNRAS.421.1838B,2010MNRAS.405.1491M}.
Therefore, we use the \textsf{GAMMIE} cooling, which should be understood as an effective cooling, which results from cooling and heating processes.\\
\cite{2001ApJ...553..174G} was able to show that the critical cooling parameter $b_\mathrm{cool}\approx 3$ generates a graviturbulent state\refree{, which can transport angular momentum very efficiently \citep{2017arXiv170402193K}. This settles in a marginally stable state with a Toomre parameter}
\begin{equation}
Q = \frac{c_\mathrm{s} \kappa}{\pi G \Sigma} \approx 1
,\end{equation}
\refree{where $\kappa$ is the epicyclic frequency}. For smaller cooling parameters the disk fragments, which could stop the accretion process onto the central object.\LEt{ There is no verb here. Perhaps: For smaller cooling parameters, the disk fragments  could stop the accretion process onto the central object.  }
Therefore, we choose a few typical values for $b_\mathrm{cool}\in\left\{1,3,6,10\right\}$ in the vicinity of the critical value.\\
The cooling timescale depends on the local dynamical timescale (see ~\eqref{eq:dynscales}, $t_\mathrm{dyn} = \sfrac{2\pi}{\Omega} = 2\pi\cdot \tau_\mathrm{dyn}$) and therefore is the range 
\begin{equation}
\tau_\mathrm{cool}\in b_\mathrm{cool}\cdot\unit[{[0.04,80]}]{Myr}.
\end{equation}
Thus, the weakest cooling of all simulations runs on a timescale of $\unit[800]{Myr}$.
In the case of the weakest cooling, we still reach two cooling timescales at the outer boundary until the simulation stops.

Furthermore, we define a power law for the temperature with a maximum near the inner boundary in simulations with an energy equation similar to the initial surface density distribution. The power law is given by
\begin{equation}
  T(r) = T_0 \cdot\min\left(\left(\frac{r}{R_\sigma}\right)^\sigma,1\right)
\end{equation}
using $\sigma=-1$ and $R_\sigma=\unit[100]{pc}$.
The solutions are \deleted{especially}  insensitive to the temperature \added{especially} in the inner region since there the cooling runs on very short timescales and an equilibrium state is reached very quickly.
In addition, the simulation should start with a rather high temperature so that the disk is Toomre stable,  otherwise a unrealistic equilibrium state could be achieved as a direct consequence of the initial conditions.\\
Accretion disks usually present highly supersonic rotation,  \LEt{ yes? } hence the kinetic energy dominates strongly over the internal energy.
Despite the conservation of total energy, a consequent transfer of  much internal to kinetic energy is possible, which means that the internal energy would become negative. To prevent this, we define a lower limit for the pressure so it has to remain positive. If the pressure falls below this limit, it is raised in these cells to its lower limit $p_\mathrm{min}$
\begin{equation}
p_{i,j} = \max\left(p_{i,j},p_\mathrm{min}\right).
\end{equation}
The local pressure minimum is set by means of a global lower temperature limit $T_\mathrm{min}$, for example  the cosmic microwave background temperature
\begin{equation}
p_\mathrm{min} = \rho_{i,j}\frac{R_\mathrm{G}}{\mu} T_\mathrm{min}.
\end{equation}
While analyzing the simulations the areas at a minimum temperature can be identified.
This is usually only the case for very short time spans whenever highly dynamical events happen, for example a strong accretion event. 
Furthermore, rearward of the shocks small extremely cold regions can be generated that are reheated by their surroundings on short timescales.\\

\section{Simulations with constant speed of sound}
In this work we focus on the impact of large-scale gas flows within a barred galaxy on the growth rate of its central black hole. The parameter space consisting of initial surface density and sound velocity is described in table~\ref{tab:agndiskisoflat}. The four \refree{values for the} speed of sound can be associated with gas temperatures of $T \in \unit[\{72,   652,  1811 ,  7244\}]{K}$ according to \eqref{eq:temp}.\\
\begin{table*}\centering
        \caption{Summary of the mean and maximum accretion rates, structure formations (gravitational instabilities, GI\LEt{ this should be in a  note below the table; also Table 3 (note 10 seems to have disappeared)}), and final black hole masses for isothermal simulations.}
        \begin{stabular}{ccc|ccccc}\hline\hline
                ID    &         $\Sigma_0$      & $c_\mathrm{s}$ & $\langle \dot{M} \rangle$ &    $\max(\dot{M})$    & Ring & Inner Disk &  $M_\mathrm{BH}$   \\
                &               {[$\unit{M_\odot/pc^2}$]} & [$\unit{km/s}$]  &   [$\unit{M_\odot/yr}$]   & [$\unit{M_\odot/yr}$] &      &                & [$\unit{M_\odot}$] \\ \hline
                s10c1\phantom{0}        &       $10$            &      $1$       &      $<10^{-2}$      &        $0.02$         &  Yes &       GI       &  $1.25\cdot 10^6$  \\
                s10c3\phantom{0} &      $10$            &      $3$       &      $<10^{-2}$      &        $0.02$         &  Yes &       GI       &  $2.44\cdot 10^6$  \\
                s10c5\phantom{0}        &       $10$            &      $5$       &      $<10^{-2}$      &        $0.02$         &  Yes &    Spiral      &  $2.58\cdot 10^6$  \\
                s10c10 &                $10$            &      $10$      &      $<10^{-2}$      &        $0.04$         &  Yes &      Shock     &  $6.73\cdot 10^6$  \\ \hline
                s30c1\phantom{0} &              $30$            &      $1$       &      $0.07 \pm 0.26$      &        $0.26$         & No   &       GI       &  $1.23\cdot 10^8$  \\
                s30c3\phantom{0} &              $30$            &      $3$       &      $0.06 \pm 0.26$      &        $0.26$         & No   &       GI       &  $1.01\cdot 10^8$  \\
                s30c5\phantom{0} &              $30$            &      $5$       &      $0.04 \pm 0.26$      &        $0.08$         & No   &       GI       &  $6.53\cdot 10^7$  \\
                s30c10 &                $30$            &      $10$      &      $0.00 \pm 0.08$      &        $0.69$         & No   &       GI       &  $8.93\cdot 10^6$  \\ \hline
                s50c1\phantom{0} &              $50$            &      $1$       &      $0.27 \pm 0.69$      &        $0.82$         & No   &       GI       &  $3.81\cdot 10^8$  \\
                s50c3\phantom{0}        &       $50$            &      $3$       &      $0.25 \pm 0.82$      &        $0.82$         & No   &       GI       &  $4.21\cdot 10^8$  \\
                s50c5\phantom{0}        &       $50$            &      $5$       &      $0.24 \pm 0.78$      &        $0.78$         & No   &       GI       &  $4.06\cdot 10^8$  \\
                s50c10 &        $50$            &      $10$      &      $0.14 \pm 0.76$      &        $0.76$         & No   &       GI       &  $2.43\cdot 10^8$\\\hline
        \end{stabular}
        \label{tab:agndiskisoflat}
\end{table*}
The generation of particular structures like nuclear rings and nuclear spirals are well described in \cite{2015ApJ...806..150L,2013ApJ...769..100S,2004MNRAS.354..892M}, and \cite{1995ApJ...449..508P}. \LEt{ When listing references in the running text of a sentence, please separate them by commas and place an “and” between the last comma and the last reference (or an “and” and no comma if there are only two references). Please check for this throughout the paper. If you have trouble with the formatting, highlight the references in red. }
\figtwovpgf{1.0}{1.0}{sims/figure/agndiskisoflat/1.E+1_4.5E-1_0.E-0_5.E+3/}{griddensitypolar_0050}{griddensitypolar_0100}{griddensitypolar_0050_0100}
{Surface density of the simulation with $\Sigma_0=\unit[10]{M_\odot/pc^2}$ and speed of sound $c_\mathrm{s}=\unit[5]{km/s}$  on different scales of the central region after a half revolution of the bar $t=\unit[93]{Myr}$ (top) and after one revolution of the bar $t=\unit[186]{Myr}$ (bottom).}
\figpgf{sims/figure/agndiskisoflat/1.E+1_4.5E-1_0.E-0_5.E+3/}{griddensitypolar_0200}{griddensitypolar_0200}{Surface density of the simulation with $\Sigma_0=\unit[10]{M_\odot/pc^2}$ and speed of sound $c_\mathrm{s}=\unit[5]{km/s}$  on different scales of the central region after two revolutions of the bar $t=\unit[372]{Myr}$.}
Nonetheless, we  introduce the typical structure formation in our galaxy model by examining the isothermal simulation with an inital surface density of $\Sigma_0=\unit[10]{M_\odot/pc^2}$ and speed of sound of $c_\mathrm{s}=\unit[5]{km/s}$;  in addition to   the typical structures, this shows a quite laminar flow.
Simulations with higher mass or lower speed of sound exhibit small spatially localized regions of significantly higher surface density than their surroundings.
These {clumps} can modify or destroy the typical structures of the gas disk.
Figures~\ref{griddensitypolar_0050_0100}--\ref{griddensitypolar_0200} show the surface density in these simulations at different zoom levels corresponding to radial scales $R\in\{\unit[16]{kpc},\unit[6]{kpc},\unit[2]{kpc},\unit[300]{pc}\}$.
In figure~\ref{griddensitypolar_0050_0100} (top) the bar potential is still raising.
The surface density shows spiral structures on the kiloparsec scale that slowly bend.
These spiral structures are a source of potential vorticity \citep{2012ApJ...747...60K} and can be subject to the wiggle instability \citep{2004MNRAS.349..270W,2014ApJ...789...68K}.
However the spatial resolution in this part of the computational domain is too low to resolve the instability at its early stages.
The performed simulations only show the instability at smaller radii, which  are reached  shortly before the gas flows from the spiral structures into the nuclear ring region.\\
In figure~\ref{griddensitypolar_0050_0100} (bottom) the bar potential has reached its full strength. Now the spiral structures incorporate a nuclear ring with semi-axes of $\unit[2.5]{kpc}$ and $\unit[2]{kpc}$.
The ring often decays into individual clumps, since the high surface density material gains high vorticity from the spiral arms.
If self-gravitation is disregarded but high spatial resolution is used, we often find clumps, which are generated by the wiggle instability at the spiral structures and are then  transported to the ring region.
The spiral structure changes its shape from slightly curved to nearly straight with a small angle inclined to the x-axis after the ring is generated.
These structures, often called off-axis shock \citep{2012ApJ...747...60K}, connect the ring on a radius of $\unit[\sim 1]{kpc}$ with the region beyond $\unit[4]{kpc}$.
The gas outside the nuclear ring typically follows strong eccentric orbits.
If it encounters one of the two off-axis shocks, it loses  most of its rotational energy and descends to a smaller orbit or can be directly dragged by other material onto the nuclear ring (see \citealp{2012ApJ...747...60K}).
Thus the ring is fed with new matter.
Furthermore, on large scales \added{as well as inside the nuclear ring }spiral arms can
be identified\deleted{ inside the nuclear ring}.\\
In figure~\ref{griddensitypolar_0200} the surface density of the nuclear ring has increased so much that has become gravitationally unstable\replaced{. At the same time}{ and} its eccentricity has decreased.
The whole structure is shaped elliptically if it rotates counterclockwise.
Inside  the nuclear ring  another ring-like  (or   disk-like) structure is recognizable.
This again has a bar-like region of low surface density and two spiral arms.\\
Numerous observations show nuclear rings (\citealp{2010MNRAS.402.2462C,2015ApJ...806L..34F,2016MNRAS.455.2745K,2016arXiv160305405K}), off-axis shocks (\citealp{2003ApJS..146..353M,2003ApJ...589..774M,2005AJ....130.1472P,2010ApJ...723..767V}), and inner spirals and bars (\citealp{2004A&A...415..941E}).
Their formation is reproduced by comparable numerical examinations done by \cite{1979ApJ...233...67R,1992MNRAS.259..345A,1995ApJ...449..508P,2004MNRAS.354..892M,2012ApJ...747...60K}, and \cite{2015ApJ...806..150L}.\LEt{ see previous note }
A detailed comparison of emerging structures with observations is given in section \ref{sec:diskussioniso}.
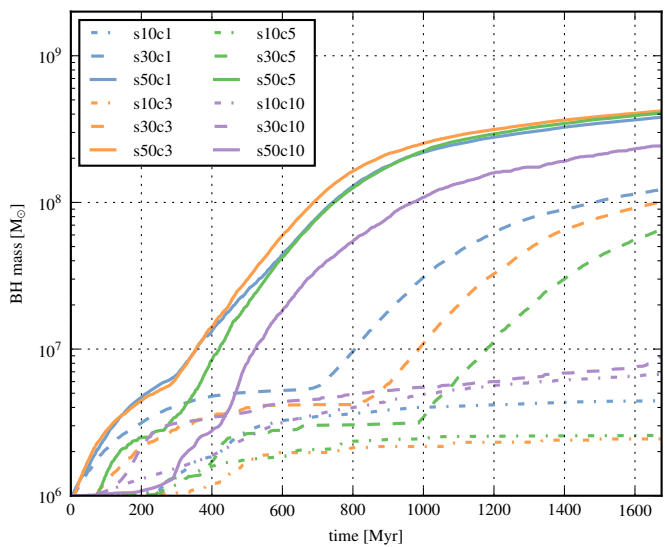
\figpgf{sims/figure/agndiskisoflat/}{bhmass_0}{isobhmass_0}{Evolution of the black hole masses of all different parameter combinations in the isothermal simulations. We note that there are two very different accretion modes. The simulations with $\Sigma_0=\unit[30]{M_\odot/pc^2}$ show  that the lower the speed of sound, the earlier the simulation switches to the strong accretion mode and the higher  the final black hole mass is.}

An overview of all the simulations with constant speed of sound can be achieved by examining the mass of the central black hole as a function of time (figure~\ref{isobhmass_0}).
Most of the simulations reach considerable black hole masses of up to $\unit[4\cdot 10^8]{M_\odot}$.
All the simulations show a switch-on phase while the bar potential is activated, and reach one of two different states at about $\unit[370]{Myr}$:
\begin{enumerate}
        \item Simulations with a high speed of sound and low disk mass accrete slowly (s10*, s30c10);
        \item Simulations with a low speed of sound and high disk mass accrete quickly (s30c1--s30c5, s50c*).
\end{enumerate}
Though the transition of simulations from slow to fast accretion modes is clearly possible, the launch of the strong accretion mode is favored by high disk masses and a  low speed of sound, as shown by the $\Sigma_0=\unit[30]{M_\odot pc^{-2}}$ and $c_\mathrm{s}=\unit[\{1,3,5\}]{km/s}$ simulations (figure~\ref{isobhmass_0}). At the end of simulations with strong accretion the black holes grow with an approximately constant accretion, as will be shown later.

\subsection{Weak accretion}
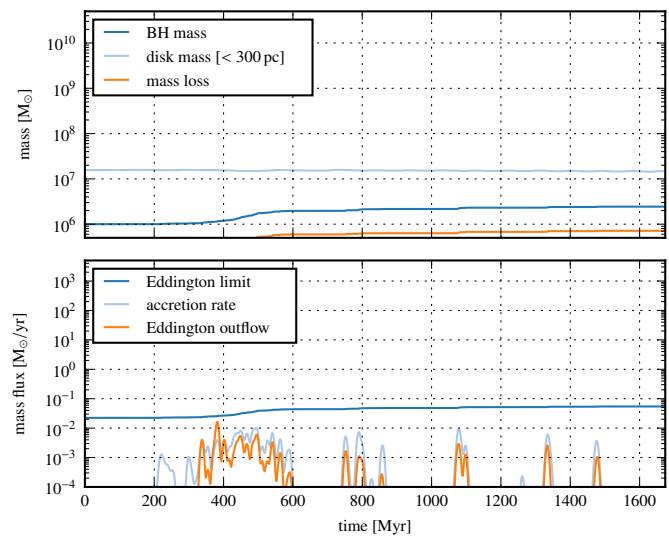
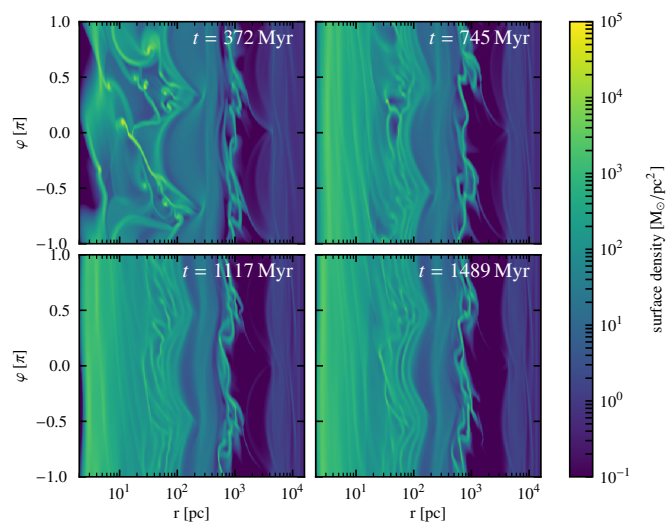
\figtwovpgf{1.0}{1.0}{sims/figure/agndiskisoflat/1.E+1_4.5E-1_0.E-0_3.E+3/}{accdiskmass}{griddensity_0200_0400_0600_0800}{10_3_accdiskmass_griddensity}{Time evolution of the simulation with $\Sigma_0=\unit[10]{M_\odot/pc^2}$ and speed of sound $c_\mathrm{s}=\unit[3]{km/s,}$  as an example of weak accretion. \\
        \textit{Top}: After the bar is switched on the accretion rate rises for a short period. Afterward it declines and falls below $\unit[10^{-4}]{M_\odot/yr}$.\\
        \textit{Bottom}: At the beginning of the simulation the surface density shows fragmentation in the inner region. Later the flow in the inner disk is laminar. Eccentric nuclear rings can be identified at $\unit[80]{pc}$ and $\unit[800]{pc}$.}
The simulation with the surface density parameter $\Sigma_0=\unit[10]{M_\odot/pc^2}$ and $c_\mathrm{s}=\unit[3]{km/s}$ is an excellent example of the weak accretion case.
Figure~\ref{10_3_accdiskmass_griddensity} (top) shows the black hole growth and the accretion rate as a function of time.
The accretion rate never exceeds $\unit[10^{-2}]{M_\odot/yr}$ and the Eddington outflow is on the same order.
The bulk of the accreted mass arises from a short accretion phase after approximately $\unit[300]{Myr}$.
At this time figure~\ref{10_3_accdiskmass_griddensity} (bottom) shows clump formation.
At later times the flow is predominantly laminar and two nuclear rings can be identified.
In particular, the low-mass gap at around $\unit[300]{pc}$ seems to prohibit further accretion since no matter can flow into the inner disk ($r<\unit[300]{pc}$).
Therefore, the inner disk remains stable and more accretion is suppressed.\\
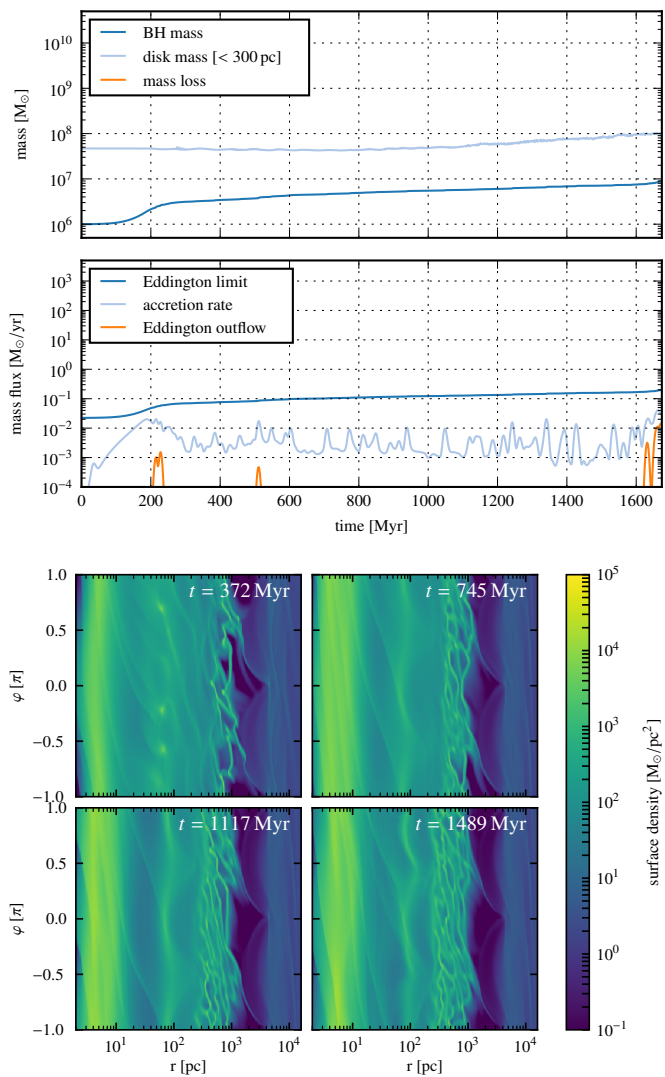
\figtwovpgf{1.0}{1.0}{sims/figure/agndiskisoflat/3.E+1_4.5E-1_0.E-0_1.E+4/}{accdiskmass}{griddensity_0200_0400_0600_0800}{30_10_accdiskmass_griddensity}{Time evolution of the simulation with $\Sigma_0=\unit[30]{M_\odot/pc^2}$ and speed of sound $c_\mathrm{s}=\unit[10]{km/s,}$  as an example of weak accretion with central shocks.\\
        \textit{Top}: Simulation showing a steady accretion rate of $\unit[10^{-3\dots -2}]{M_\odot/yr}$.\\
        \textit{Bottom}: First fragmentation occurring around $\unit[100]{pc}$ and inside the nuclear ring at $\unit[800]{pc}$. Later on the inner fragmentation fades and a rather eccentric ring and inner shocks  form at $\unit[80]{pc}$.}
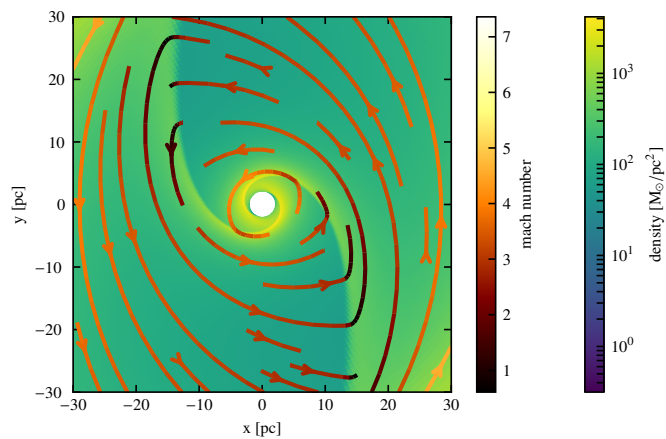
\figpgf{sims/figure/agndiskisoflat/3.E+1_4.5E-1_0.E-0_1.E+4/}{streamline_density_3.0e+01_0100}{30_10_stream_0100}{Instantaneous streamlines and surface density map of the simulation with $\Sigma_0=\unit[30]{M_\odot/pc^2}$ and speed of sound $c_\mathrm{s}=\unit[10]{km/s}$.  At the shocks the gas is compressed and loses a great part of its kinetic energy. Its velocity becomes subsonic and the gas is pushed to lower radii until it again reaches sufficient velocity to balance centrifugal and gravitational forces. This process can drive the gas  far into the inner disk within a few orbits.}
Figure~\ref{30_10_accdiskmass_griddensity} illustrates the time evolution of the
simulation with higher initial mass ($\Sigma_0=\unit[30]{M_\odot/pc^2}$) and greater speed of sound ($c_\mathrm{s}=\unit[10]{km/s}$).
The mass of the nuclear ring at $\unit[\sim 800]{pc}$ is sufficient to refill the inner disk after its initial mass loss.
However, the  relevant process here is not clumpy accretion, but loss of kinetic energy by shock compression (e.g., \citealp{1992aita.book.....S}).
Since this is an isothermal simulation, the gas temperature stays constant.
In order to illustrate how the gas moves in the inner areas, figure~\ref{30_10_stream_0100} depicts the velocity field using instantaneous streamlines.
When passing through a shock, a substantial fraction of the kinetic energy is dissipated leading to a decelaration along the azimuthal direction. Hence, gravitational forces are no longer balanced by centrifugal forces and therefore the gas experiences a net radial force which causes a radial inflow. It should be noted that the dissipated energy is lost because the temperature is kept constant in the isothermal simulations. The accretion rate onto the black hole (figure~\ref{30_10_accdiskmass_griddensity}) is  in the range $\unit[10^{-3\dots -2}]{M_\odot/yr}$ without large deviations for more than $90\%$ of
the time.\LEt{ have I interpreted correctly? }\\
It should be noted that the accretion rate and Eddington outflow rate are mean values taken over the period between two data outputs and can therefore not resolve fluctuations, which are shorter than $\Delta t=\unit[1.86]{Myr}$.
If the accretion rate exceeds the Eddington limit even for a short period $<\unit[1.86]{Myr}$, outflow is generated, although the mean accretion rate never reaches the Eddington limit (see  figure~\ref{30_10_accdiskmass_griddensity}).
Thus, in this simulation less than $\unit[\sim 5\cdot 10^{5}]{M_\odot}$ is lost by means of the Eddington limit because the accretion rate is almost constant and the Eddington outflow rate is usually below $\unit[10^{-4}]{M_\odot/yr}$.
Shortly before the end of the simulation, the inner disk reaches a mass of about $\unit[10^8]{M_\odot}$, which is  the requirement for strong accretion, as is discussed in the next section.

\subsection{Strong accretion}
First we consider simulations, where the strong accretion starts at a later point in time after the switch-on process is finished. Figure~\ref{30_3_accdiskmass_griddensity} (top) shows the time evolution of accretion rate, Eddington outflow rate, and black hole mass for the simulation with $\Sigma_0=\unit[30]{M_\odot/pc^2}$, $c_\mathrm{s}=\unit[3]{km/s}$. During the initial phase of about $\unit[800]{Myr,}$ the accretion rate and the outflow rate are roughly on the same order and decrease from $\unit[10^{-2}]{M_\odot/pc^2}$ to below $\unit[10^{-4}]{M_\odot/pc^2}$. In this phase we observe strong variations which indicate clumpy accretion. From time to time individual clumps cross the inner boundary leading to accretion rates above the Eddington limit for short periods. This does not show up in the diagram because the outlined accretion rate sums the accretion between two data outputs over a period of $\unit[1.86]{Myr}$.
Therefore, short-lived accretion events cannot be resolved.
Figure~\ref{30_3_accdiskmass_griddensity} (bottom) shows the surface density for four selected points in time.
The figure for $\unit[372]{Myr}$ supports the observation of a clumpy accretion, given that multiple clumps are visible in the inner disk.
The free-fall timescale of a clump at  radius $R=\unit[30]{pc}$ for the initial black hole mass $M=\unit[10^6]{M_\odot}$ is $\tau_\mathrm{ff} = \unit[2.45]{Myr}$.
This is comparable to the time resolution of the simulations for data outputs of $\unit[1.86]{Myr}$.
Thus, it cannot be expected to directly observe the infall of a clump.\\
This first clumpy accretion phase is completed after $\unit[745]{Myr}$ since the flow has adapted to the fully switched on bar potential and there is no further accretion.
In the inner disk we find three round rings (figure~\ref{30_3_accdiskmass_griddensity}, bottom, vertical lines in this projection).
Initially the mass of the inner disk ($\unit[<300]{pc}$) is reduced by approximately $\unit[~5\cdot 10^6]{M_\odot}$.
Starting at the region around $\unit[1]{kpc}$ the accretion disk becomes gravitationally unstable \citep{1960AnAp...23..979S,1964ApJ...139.1217T,1978AcA....28...91P,1979AcA....29..157K,2015ApJ...804...62R}, which drives an accretion flow inward. Therefore the inner regions start gaining mass and the unstable region expands inward. At $\unit[400]{Myr}$ the inner disk mass ($r<\unit[300]{pc}$) starts increasing.
If the mass in the inner disk reaches $\unit[10^8]{M_\odot}$, it becomes gravitationally unstable and the strong accretion mode, which is initially very clumpy, sets in.
This kind of gravitational instability, which generates individual gravitationally bound structures, is typical for systems on very short cooling timescales \citep{2001ApJ...553..174G}.
In this case an isothermal equation of state is used, which is equivalent to an instantaneous cooling, since the gas cannot be heated by compression.
The Toomre parameter \citep{1964ApJ...139.1217T} characterizes whether gravitational instabilities can be expected in an accretion disk.
The parameter is proportional to $Q\propto \sqrt{T}/\Sigma$. Since the temperature is constant in this case, a stable state $Q>1$ can only be reached again by decreasing the surface density $\Sigma$.
If the accretion disk becomes Toomre unstable by accumulation of mass, gravitational instabilities arise.\\
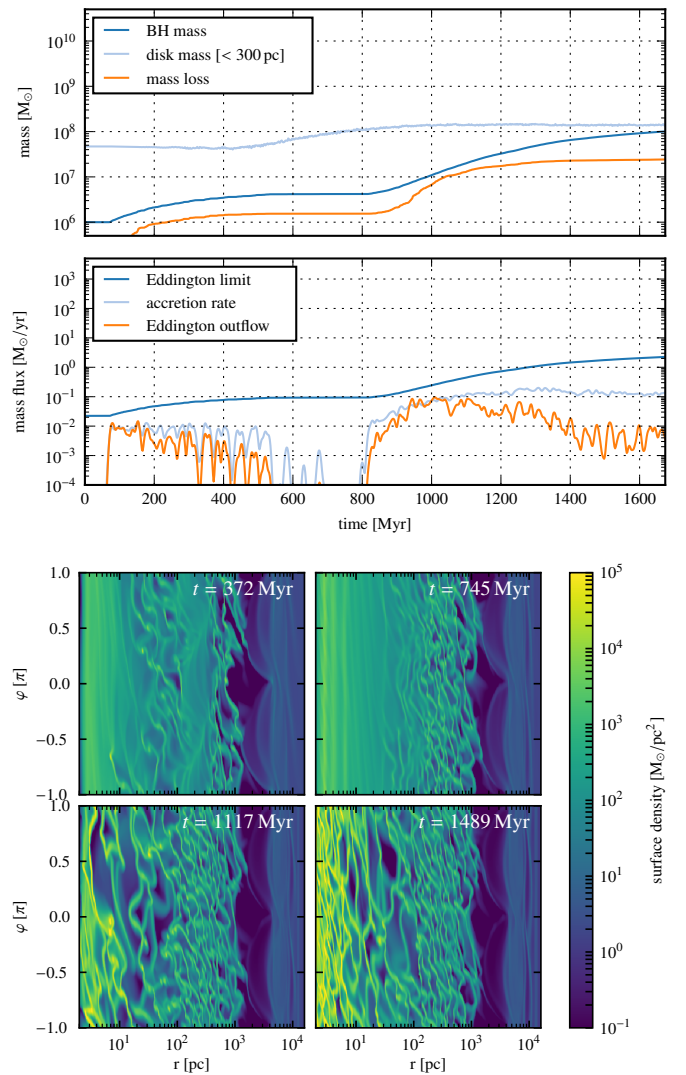
\figtwovpgf{1.0}{1.0}{sims/figure/agndiskisoflat/3.E+1_4.5E-1_0.E-0_3.E+3/}{accdiskmass}{griddensity_0200_0400_0600_0800}{30_3_accdiskmass_griddensity}{Time evolution of the simulation with $\Sigma_0=\unit[30]{M_\odot/pc^2}$ and speed of sound $c_\mathrm{s}=\unit[3]{km/s,}$  as an example of strong accretion.\\
        \textit{Top}: After a half revolution of the bar, a strong clumpy accretion flow with high Eddington outflow already starts, and decreases with time. Later on the strong accretion is established, which again is highly clumpy at first and reaches an accretion rate of $\unit[0.1]{M_\odot/yr}$.\\
        \textit{Bottom}: Initial clump formation  is observed in the inner disk, but  fades quickly. Then the inner disk becomes laminar until the strong accretion sets in. Afterward, the complete disk inside  $\unit[1]{kpc}$ exhibits strong clump formation before a gravitational turbulent mode is reached.}
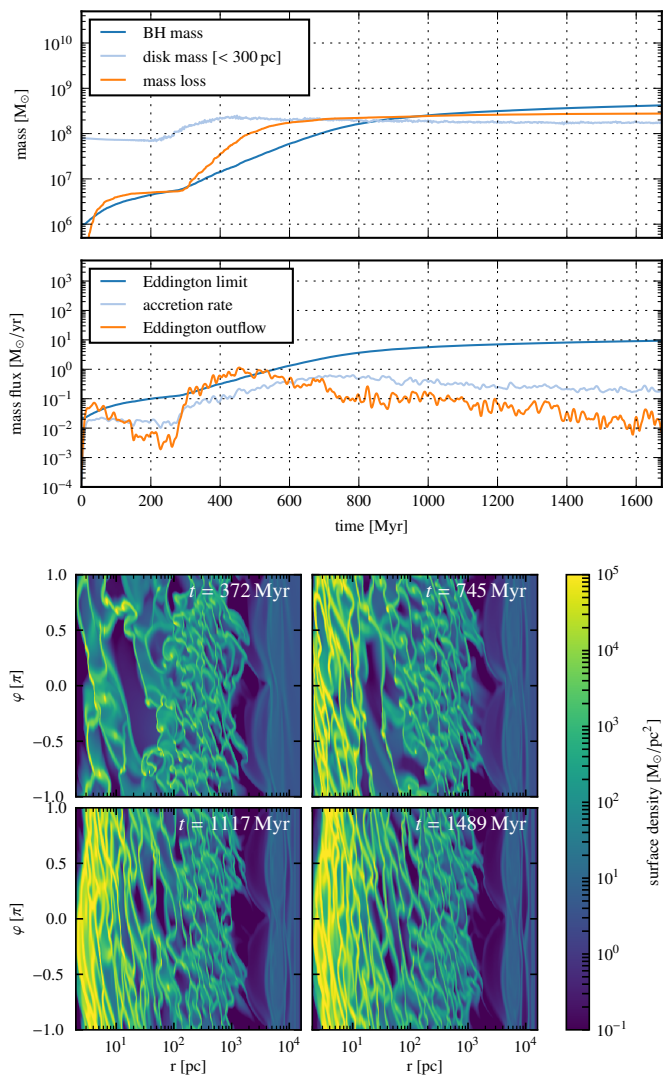
\figtwovpgf{1.0}{1.0}{sims/figure/agndiskisoflat/5.E+1_4.5E-1_0.E-0_3.E+3/}{accdiskmass}{griddensity_0200_0400_0600_0800}{50_3_accdiskmass_griddensity}{Time evolution of the simulation with $\Sigma_0=\unit[50]{M_\odot/pc^2}$ and speed of sound $c_\mathrm{s}=\unit[3]{km/s,}$  as an example of strong accretion.\\
        \textit{Top}: Simulation reaching the highest final black hole mass of all the simulations carried out for this work. During the switch-on phase an inner disk mass of more than $\unit[10^8]{M_\odot}$ is reached. This leads immediately to strong accretion of more than $\unit[0.1]{M_\odot/yr}$ right from the beginning.\\
        \textit{Bottom}: Simulation initially exhibiting strong fragmentation, before  changing to a gravitational turbulent accretion mode. A laminar inner disk is never observed.}
Figure~\ref{30_3_accdiskmass_griddensity} (bottom) already shows the generation of instabilities at a radius of $\unit[800]{pc}$   at $\unit[372]{Myr}$, which spread until $\unit[1117]{Myr}$ to smaller and smaller radii, and therefore lead to an increase in the inner disk mass.
After about $\unit[1100]{Myr}$ the accretion rate settles at $\unit[10^{-1}]{M_\odot/yr}$ and the outflow rate declines.
From this time on the inner disk mass stays approximately constant at $\unit[1.5\cdot 10^8]{M_\odot}$.
The inner disk is supplied with mass from the outer disk at the same rate as mass flows over the inner boundary, which is then  accreted by the black hole.
The Eddington outflow is about one order of magnitude below the accretion rate at the end of the simulation.
An examination of the surface density distribution reveals that the accretion is less clumpy than before. Instead,  clumps are so numerous in the central disk $\unit[<10]{pc}$ that they are sheared out to density filaments. The accretion flow is still equally strong, but smoother and less abrupt, and there are three different epochs:
\begin{description}
        \item[{\unit[0--800]{Myr}},] where an initial accretion flow of $\unit[10^{-4\ldots-2}]{M_\odot/yr}$ sources its matter from the inner disk and refills it from outside;
        \item[{\unit[800--1100]{Myr}},] where the inner disk is still being filled and drives a strong clumpy accretion flow $\unit[10^{-2\ldots -1}]{M_\odot/yr}$;
        \item[{\unit[1100--1600]{Myr}},] where strong accretion with a constant rate of $\unit[10^{-1}]{M_\odot/yr}$ takes place and the inner disk is refilled from outside at the same rate.
\end{description}
The simulation with initial surface surface density $\Sigma_0=\unit[50]{M_\odot/pc^2}$ and speed of sound $c_\mathrm{s}=\unit[3]{km/s}$  is the isothermal simulation with the highest black hole mass at the end of the simulation time.
The simulation experiences a switch-on phase, which directly leads into a strong accretion mode since an inner disk mass of $\unit[>10^8]{M_\odot}$ is already reached after $\unit[300]{Myr}$.
Figure ~\ref{50_3_accdiskmass_griddensity} shows an initially clumpy accretion with strong Eddington outflow.
Afterward, up to a radius of about $\unit[100]{pc}$, \added{mainly} filamentary structures can \deleted{mainly}  be identified.
At $\unit[1000]{Myr}$ the accretion rate settles down to $\unit[2\cdot 10^{-1}]{M_\odot/yr}$,  and the Eddington outflow decreases. In addition,  the unstable region between $\unit[100]{pc}$ and $\unit[1]{kpc}$ no longer shows clumps; \LEt{ or: does not show any additional clumps.} instead, this region is dominated by a filamentary flow.
\subsection{Time evolution of black hole masses}\label{sec:accandagnev}
Figure~\ref{isobhmass_0} shows that cold disks with higher initial mass (e.g., $c_\mathrm{s}=\unit[3]{km/s}$ and $\Sigma_0=\unit[50]{M_\odot/yr}$) induce faster accretion, which then leads to higher black hole masses at the end of the simulations.
Table~\ref{tab:agndiskisoflat} outlines the mean accretion rates and the final black hole masses. Again, this indicates that massive gas disks accrete more strongly.
For simulations with $\Sigma_0=\unit[10]{M_\odot/pc^2}$ the maximum accretion rate is $\unit[0.02]{M_\odot/yr}$ and the temporal mean is even below $\unit[<10^{-2}]{M_\odot/yr}$.
Nuclear rings can only persist in the least massive simulations, and even here they are clumpy because of the high surface densities.
In the massive simulations the rings are so dense that they fragment and their structures disintegrate. Furthermore, if strong accretion sets in very early, ring formation can be suppressed completely. This may only happen in simulations with the highest initial mass.
The inner disk exhibits  mostly gravitational instabilities (GI) for the isothermal simulations.
Only the mass-poor simulations with high sound velocities show the formation of  tightly wound inner spirals or strong shocks.
Clearly this is a more efficient accretion mechanism than in the simulations with $\Sigma_0=\unit[10]{M_\odot/pc^2}$.
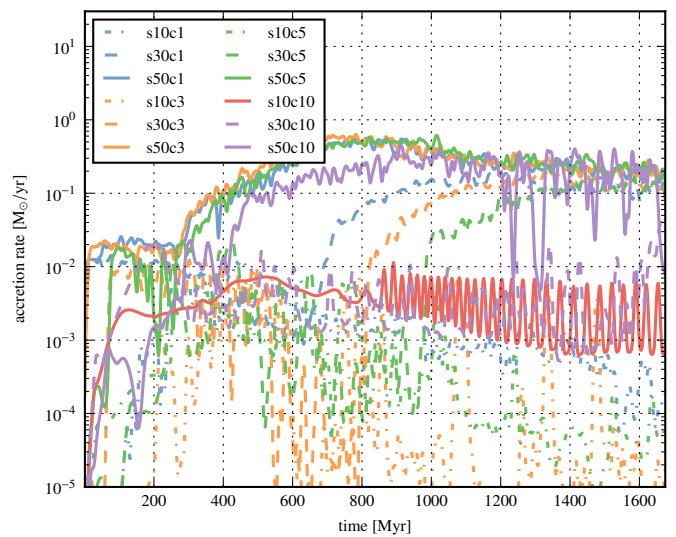
\figpgf{sims/figure/agndiskisoflat/}{accretionrate_0}{accretionrate_0_iso}{Plot of the accretion rates of all isothermal simulations. It is clearly visible that the simulations all lead   either to  strong accretion of more than $\unit[0.1]{M_\odot/yr}$ or to weak accretion of less than $\unit[0.01]{M_\odot/yr}$. Additionally, all simulations exhibit strong variations in the accretion rate.}
A variation in the sound velocities tends to result in varying final black hole masses, but more importantly influences if and when the strong accretion starts.
Figure~\ref{accretionrate_0_iso} shows the accretion rates of all isothermal simulations.
The different accretion types are clearly distinguishable.
The accretion rate is either greater than $\unit[0.1]{M_\odot/yr}$ or below $\unit[0.01]{M_\odot/yr}$.
It is striking that all the simulations show strong variability in the accretion rate.
This is probably because the inner boundary of the computational domain at $\unit[2]{pc}$ is not close enough to the black hole, and therefore an important part of the accretion disk lies outside  the computational domain.
It is still assumed that the mass that leaves the computational domain across the inner boundary is directly accreted onto the black hole.
This is especially problematic for clumpy accretion.
At small radii the differential rotation becomes stronger and stronger, hence the clumps would be sheared out to density filaments.
The accretion of filaments is much more continuous, so that the effective accretion rate onto the black hole might be closer to the Eddington limit.\\
A special case is the simulation with $\Sigma_0=\unit[10]{M_\odot/pc^2}$ and a speed of sound of $c_\mathrm{s}=\unit[10]{km/s}$.
Here from $\unit[800]{Myr}$ onward the accretion rate  oscillates  with a period of about $\unit[20-25]{Myr}$ (figure~\ref{accretionrate_0_iso}, red line).
On closer examination of the density structure at different times, the origin of the periodicity becomes clear.
In this simulation a nuclear ring develops and, since the onset of the periodicity, an inner shock becomes visible.
The eccentric nuclear ring oscillates on the above-mentioned time scales, and therefore ejects mass at regular intervals across shocks into the inner disk region and the black hole.
The oscillation period decreases slowly with time. The oscillation does not fade; instead, it remains  stable for more than $\unit[800]{Myr}$.\\

\subsection{Discussion}\label{sec:diskussioniso}
Clumpy accretion plays an important role in many of the examined simulations.
Primarily isothermal simulations with strong accretion encounter Eddington outflows that are  almost as strong as the accretion rate.
The simple implementation of the Eddington limit may not be sufficient any longer.
On the one hand, a substantial part of the accretion disk is outside  the computational domain at radii $\unit[<2]{pc}$.
Here clumps could decay to a more homogeneous disk, which transports the matter more continuously to the black hole. Also, new clumps could not be formed in the innermost part of the disk since it is not self-gravitating and therefore Toomre stable.
This would be especially important for the infall of single massive clumps.
However, if the outflow rate is uniformly very high, there would be no particular effect.
On the other hand, the Eddington limit is not a strict maximum of a variable accretion rate.
It describes the equilibrium state of the balance of gravitational forces and Thomson scattering \citep{1979rpa..book.....R,frank2002accretion}.
This means that massive clumps with strong inward directed momentum, which can be  observed after a typical clump-clump interaction for example, are slowed down by the Thomson scattering, but still reach the black hole.
However, in order to model these effects we have to resolve the flow within the inner disk on the subparsec scale.
These model limitations require the interpretation of the measured accretion rates  as lower limits especially in the case of clumpy accretion.
The accretion rate and Eddington outflow of such a simulation are shown in Figure~\ref{50_3_accdiskmass_griddensity} (top).
Nonetheless, the total effect on the final black hole mass can be neglected;  even if the Eddington limit were not applied, the black hole would gain at most a factor of $2$ in mass.
However,  an overall different evolution of the simulation and accretion would be conceivable since the mass loss by means of the Eddington limit has an influence on the mass and therefore on the relevance of self-gravitation in the inner disk.\\
Nuclear rings are observed in a multitude of systems \citep{2003ApJS..146..353M,2003ApJ...589..774M,2015ApJ...806L..34F,2016arXiv160305405K,2016MNRAS.455.2745K}, some with clearly clumpy structures (e.g., \citealp{2015ApJ...799...11X}) and even in the galactic center \citep{2013ApJ...775...37L}.
The observations feature surface densities of $\unit[10^2]{M_\odot/pc^2}$ \citep{2015ApJ...806L..34F} in the inner kiloparsec and up to $\unit[10^4]{M_\odot/pc^2}$ for the clumpy rings \citep{2015ApJ...799...11X}.
However the measurements are subject to large uncertainties, especially because of the required conversion factor from the indicator HCN to the gas density.
The clumpy ring structures are consistent with the results in this work, which show in the area around $\unit[100]{pc}$ densities of up to $\unit[10^4]{M_\odot/pc^2}$.\\
\cite{2015ApJ...806L..34F}  have determined a stability parameter $Q$ \citep{2011MNRAS.416.1191R,2013MNRAS.433.1389R} similar to the Toomre parameter for their HCN gas observation, which takes the disk thickness into consideration.
They conclude that the inner disk $\unit[<500]{pc}$ is stable and the ring region $\unit[\sim 800]{pc}$ is unstable.
The simulations in this work often exhibit  a gravitationally unstable inner disk \added{($\unit[<300]{pc}$)} for the region inside of a nuclear
ring\deleted{ ($\unit[<300]{pc}$)}.
This cannot be confirmed by the observations, and is evidence that stabilizing elements should be considered.
Possibly, an effective speed of sound that includes the stabilizing effects of small-scale turbulence should be leveraged \citep{2010MNRAS.407.1223R}.

\section{Simulations with cooling}\label{sec:tempstruct}
\begin{table*}\centering
        \caption{Summary of the mean and maximum accretion rates, structure formations (gravitational instabilities, GI), and final black hole masses for simulations with cooling.}
        \begin{stabular}{ccc|ccccc}\hline\hline
ID &            $\Sigma_0$          &     $b_\mathrm{cool}$      & $\langle \dot{M} \rangle$ &    $\max(\dot{M})$    & Ring & Inner disk &  $M_\mathrm{BH}$   \\
 &              {[$\unit{M_\odot/pc^2}$]} &    &   [$\unit{M_\odot/yr}$]   & [$\unit{M_\odot/yr}$] &      &                & [$\unit{M_\odot}$] \\ \hline
s10b1\phantom{0} &              $10$            &        $1$        &      $0.01\pm 0.02$       &        $0.21$         &  Yes &       GI       &  $1.43\cdot 10^7$  \\
s10b3\phantom{0} &              $10$            &        $3$        &      $0.01 \pm 0.02$      &        $0.22$         &  Yes &       GI       &  $1.85\cdot 10^7$  \\
s10b5\phantom{0}        &       $10$            &        $6$        &      $0.08\pm 0.12$       &        $0.66$         &  Yes &     Shock      &  $1.34\cdot 10^8$  \\
s10b10 &                $10$            &       $10$        &      $0.10 \pm 0.16$      &        $0.74$         &  Yes &     Shock      &  $1.68\cdot 10^8$  \\ \hline
s30b1\phantom{0} &              $30$            &        $1$        &      $0.13 \pm 0.09$      &        $0.62$         & No   &       GI       &  $1.90\cdot 10^8$  \\
s30b3\phantom{0} &              $30$            &        $3$        &      $0.11 \pm 0.23$      &        $2.25$         & No   &       GI       &  $1.86\cdot 10^8$  \\
s30b5\phantom{0} &              $30$            &        $6$        &      $0.45 \pm 0.46$      &        $3.14$         & No   &     Shock      &  $7.51\cdot 10^8$  \\
s30b10 &                $30$            &       $10$        &      $0.46 \pm 0.39$      &        $2.22$         & No   &     Shock      &  $7.70\cdot 10^8$  \\ \hline
s50b1\phantom{0} &              $50$            &        $1$        &      $0.24 \pm 0.23$      &        $2.13$         & No   &       GI       &  $4.03\cdot 10^8$  \\
s50b3\phantom{0} &              $50$            &        $3$        &      $0.51 \pm 0.65$      &        $3.16$         & No   &   GI/Shock     &  $8.59\cdot 10^8$  \\
s50b5\phantom{0} &              $50$            &        $6$        &      $0.73\pm 0.72$       &        $4.95$         & No   &     Shock      &  $1.23\cdot 10^9$  \\
s50b10 &                $50$            &       $10$        &      $0.75 \pm 0.60$      &        $3.34$         & No   &     Shock      &  $1.27\cdot 10^9$
        \end{stabular}
        \label{tab:agndiskflat}
\end{table*}
The results from the simulations with cooling are considerably more complex than the results from the isothermal simulations. 
Several different solutions for the surface density and temperature structure exist, which affect the accretion onto the black hole significantly.
The examined parameter space consists of the initial surface density and the dimensionless cooling parameter, and is shown in table~\ref{tab:agndiskflat}.\\
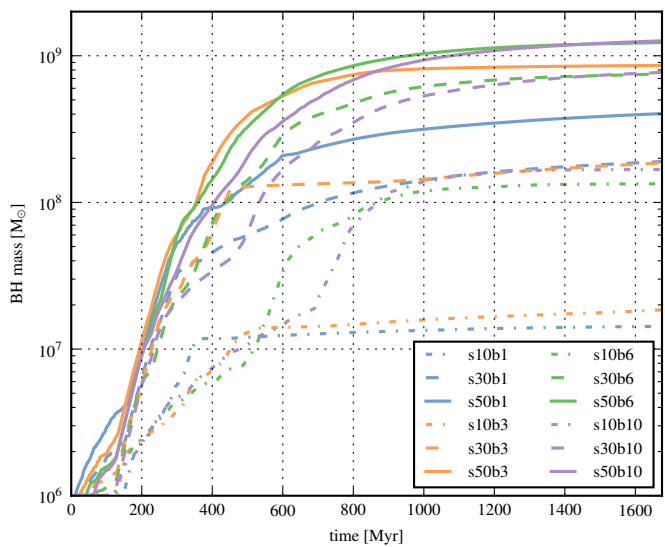
\figpgf{sims/figure/agndiskflat/}{bhmass_0}{bhmass_0}{Summary of the black hole growth for the complete parameter study of the galaxy simulations with cooling. All the simulations accrete at the Eddington limit; the accretion then decreases heavily for simulations with strong cooling.}
First, we look at the black hole masses of all simulations as a function of time in figure~\ref{bhmass_0}.
A few simulations start directly accreting as a consequence of the initial conditions.
After about one rotation of the bar all simulations accrete near the Eddington limit.
\replaced{The trigger seems to affect the   simulations that accrete below this limit, which  now also  accrete at the Eddington limit.}{The trigger seems to affect the beforehand accreting simulations as well, which now also accrete at the Eddington limit.} \LEt{
Perhaps something like this: The trigger also seems to affect the   simulations that accrete below this limit, but  now they too  accrete at the Eddington limit. }\\
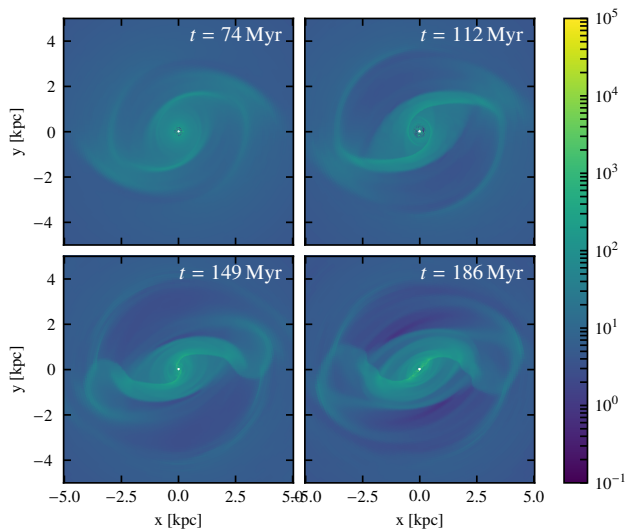
\figpgf{sims/figure/agndiskflat/3.E+1_4.5E-1_0.E-0_1.E+0/}{griddensitypolar_0040_0060_0080_0100}{schock_40_60_80_100}{Shock generation in simulations  with energy equation  similar to the isothermal simulations with $\Sigma_0=\unit[30]{M_\odot/pc^2}$ and $b_\mathrm{cool}=1$. The shocks do not connect to a ring, but  bend further until they touch the inner boundary of the computational domain. }
Figure~\ref{schock_40_60_80_100} exposes the common trigger of this very strong accretion mechanism.
Similarly to the isothermal simulations, the bar potential induces strong shocks in the gas, which grow in curvature with time.
In this case, however,  the shocks do not connect to form a ring or clumpy ring, but bend further and migrate inward until they touch the inner boundary of the computational domain.
Here they stay fixed, while the outer parts spread matter in chaotic shocks in all directions.
In the process the spiral arms heat the inner disk to very high temperatures ($\unit[10^6]{K}$).
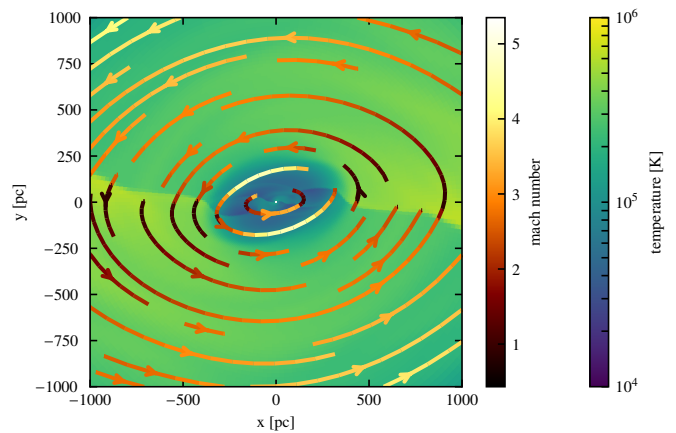
\figpgf{sims/figure/agndiskflat/3.E+1_4.5E-1_0.E-0_1.E+1/}{streamline_temp_1.0e+03_0150}{streamspiral_0150}{Strong shocks   along the bar which strongly heat the gas. The loss of kinetic energy results in a decrease in the orbits of fluid particles, which  quickly reach the central disk on spiral paths.}
The shocks are a very effective mechanism  for transporting the gas to the galaxy center \citep{1992MNRAS.259..345A}.
Figure~\ref{streamspiral_0150} shows that the gas dissipates a great deal of its kinetic energy at the shocks,  therefore heats its surroundings, and closes in on the black hole because of its low rotational velocity.
Thus, the orbits become extremely eccentric even in the direct vicinity of the black hole, whose gravity is already  very important.
The black hole growth shows that this enables a very effective accretion flow.\\
The accretion at the Eddington limit cannot be sustained after a short time by the $\Sigma_0=\unit[10]{M_\odot/pc^2}$ simulations.
Subsequently, all of them accrete more slowly until those with strong cooling abruptly diminish their accretion rates and accrete more weakly ($t_\mathrm{cool}\in\{1,3\}$ after $\unit[350]{Myr}$ and $\unit[500]{Myr, respectively}$).
We consider this cold accretion mode in more detail in the following section.
\subsection{Cold accretion}
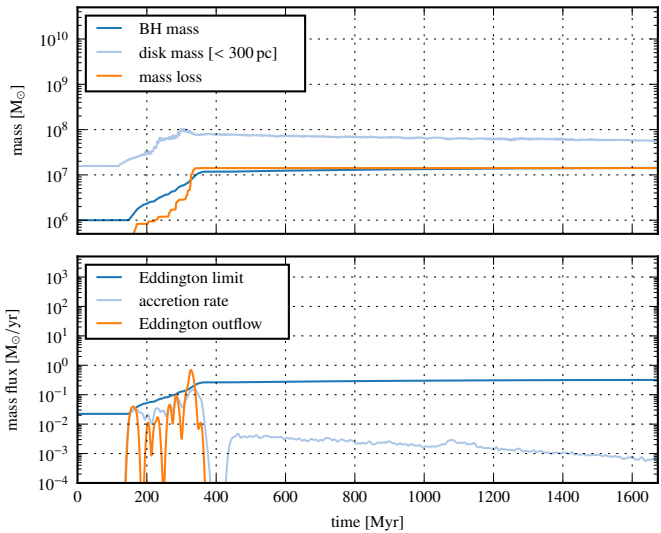
\figpgf{sims/figure/agndiskflat/1.E+1_4.5E-1_0.E-0_1.E+0/}{accdiskmass}{10_1_accdiskmass}{Time evolution of the accretion rate, the Eddington outflow, and the black hole mass of the simulation with $\Sigma_0=\unit[10]{M_\odot/pc^2}$ and $b_\mathrm{cool}=1$. Disregarding of a short accretion phase after the bar is switched on, there is only weak accretion.}
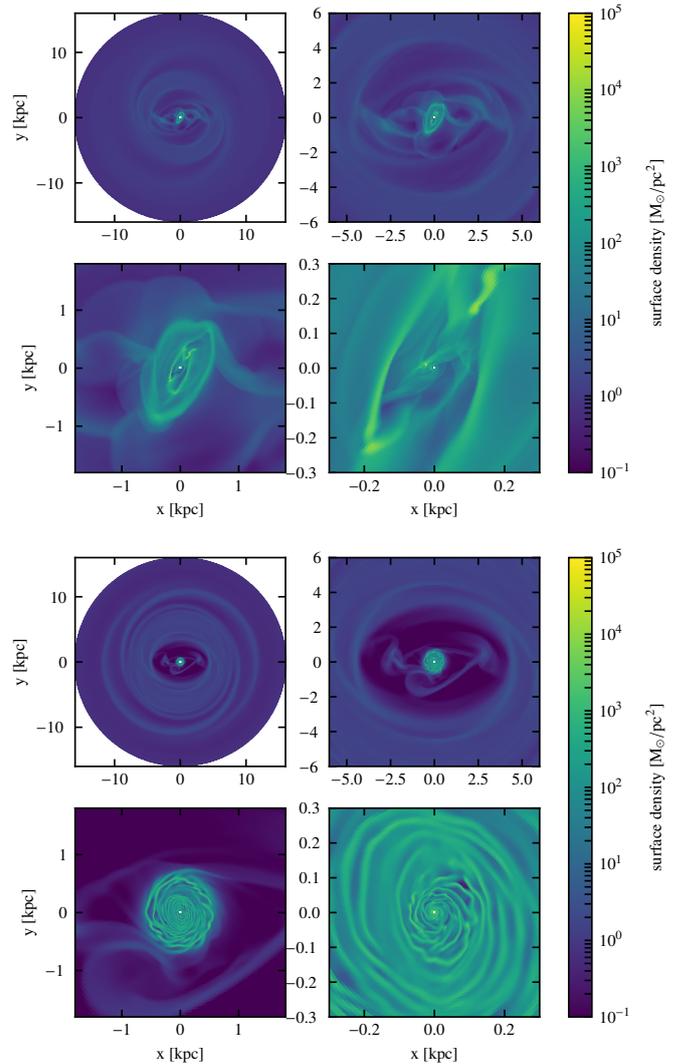
\figtwovpgf{1.0}{1.0}{sims/figure/agndiskflat/1.E+1_4.5E-1_0.E-0_1.E+0/}{griddensitypolar_0150}{griddensitypolar_0400}{kalt_griddens_0150_0400}{Surface density of the simulation with $\Sigma_0=\unit[10]{M_\odot/pc^2}$ and $b_\mathrm{cool}=1$ at different spatial scales.\\
\textit{Top} ($\unit[253]{Myr}$): In the center an irregular ring can be seen. The overall structure is very chaotic and some high-mass fragments are perceivable.\\
\textit{Bottom} ($\unit[744]{Myr}$): In the central region a marginally stable self-gravitating disk has formed, embedded in a gas-poor environment. Thus, the central accretion disk couples only weakly to the kiloparsec-scale gas disk. Angular momentum transfer to the outer parts of the galaxy is suppressed and consequently the accretion rate onto the black hole declines.}
An example of cold accretion is the simulation with $\Sigma_0=\unit[10]{M_\odot/pc^2}$ and cooling parameter $b_\mathrm{cool}=1$.
Figure~\ref{10_1_accdiskmass} shows the accretion rate and the Eddington mass loss as a function of time.
After the strong accretion phase, which all the simulations experience  initially, the accretion rate weakens in this simulation past $\unit[370]{Myr}$ very quickly.
At the time when the accretion rate has completely collapsed, a steady accretion flow of initially $\unit[4\cdot 10^{-3}]{M_\odot/yr}$ starts, which decreases exponentially  with time.
A mass loss by means of the Eddington limit as has been observed for strong accretion does not occur.
Even so,  within a short period more than $\unit[10^7]{M_\odot}$ is lost, which is approximately the same amount the black hole has accreted.
The most important feature of this plot is the drop in accretion rate shortly after the mass of the inner disk starts decreasing. 
Prior to this a fast growth phase takes place, which is completed after about $\unit[300]{Myr}$.
This is probably a switch-on effect, which starts at $t=\unit[186]{Myr}$ when the bar has reached its full strength.
There are no indications that this process might be repeated at a later time.
The inner disk loses more mass than is accreted by the black hole or than is removed by the Eddington limit from the system.
Therefore, the inner disk transfers mass to the outer regions of the galaxy.\\
Figure~\ref{kalt_griddens_0150_0400} shows the surface density structure during the strong accretion and after the drop in accretion rate.
In the strong accretion phase the flow in the kiloparsec region is comparatively chaotic.
Ring-like structures are indeed present, but they are quite amorphous, and a large amount of gas flows along the bar onto these central structures.\\

At the time when the accretion rate shows a considerable decline, a marginally stable self-gravitating disk forms in the center.
This means that gravitational instabilities arise if an accretion disk becomes Toomre unstable by cooling or accumulation of mass. These instabilities dissipate energy and heat the accretion disk.
Since the cooling is weaker than the heating by the instabilities, the Toomre parameter increases to $Q>1$ and the disk becomes stable again.
Given that the cooling still operates, this process will start over again.
Therefore, a marginally stable state  is reached with a Toomre parameter of $Q\sim 1$, as has been predicted by \cite{1978AcA....28...91P}.
This can be achieved in non-isothermal systems with slow cooling \citep{1978AcA....28...91P,2003MNRAS.339.1025R,2005MNRAS.364L..56R,2007prpl.conf..607D,2009MNRAS.393.1157C,2010MNRAS.401.2587C}. Therefore further accretion would be expected as a result of the marginally stable accretion disk.\\
Given that the region outside  the accretion disk is very gas poor, no effective coupling with the disk can be established. Angular momentum cannot be transported from the accretion disk to outside regions (see \citealp{2015MNRAS.450..691I}).
\begin{figure*}
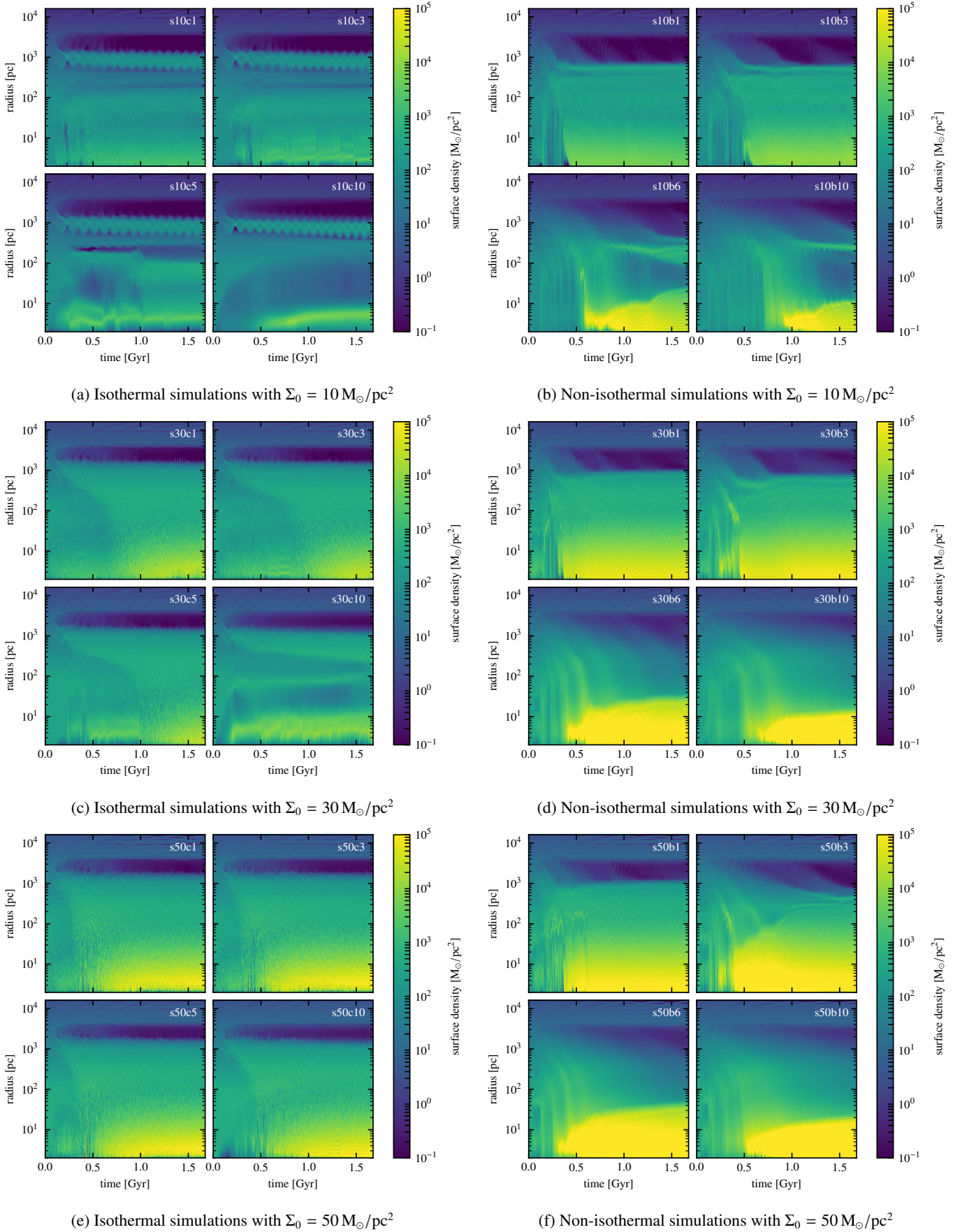
\centering
        \begin{subfigure}[b]{0.49\textwidth}
                
\renewcommand\pgfimage[2][NOSTD]{\includegraphics[#1.0]{sims/figure/agndiskisoflat/#sims/figure/agndiskisoflat/}}\resizebox{1.0\linewidth}{!}{\input{sims/figure/agndiskisoflat/densityradius_5_0.pgf}}
                \caption{Isothermal simulations with $\Sigma_0=\unit[10]{M_\odot/pc^2}$}
        \end{subfigure}
        \begin{subfigure}[b]{0.49\textwidth}
                
\renewcommand\pgfimage[2][NOSTD]{\includegraphics[#1.0]{sims/figure/agndiskflat/#sims/figure/agndiskflat/}}\resizebox{1.0\linewidth}{!}{\input{sims/figure/agndiskflat/densityradius_5_0.pgf}}
                \caption{Non-isothermal simulations with $\Sigma_0=\unit[10]{M_\odot/pc^2}$}
        \end{subfigure}\\
        \begin{subfigure}[b]{0.49\textwidth}
                
\renewcommand\pgfimage[2][NOSTD]{\includegraphics[#1.0]{sims/figure/agndiskisoflat/#sims/figure/agndiskisoflat/}}\resizebox{1.0\linewidth}{!}{\input{sims/figure/agndiskisoflat/densityradius_6_0.pgf}}
                \caption{Isothermal simulations with $\Sigma_0=\unit[30]{M_\odot/pc^2}$}
        \end{subfigure}
        \begin{subfigure}[b]{0.49\textwidth}
                
\renewcommand\pgfimage[2][NOSTD]{\includegraphics[#1.0]{sims/figure/agndiskflat/#sims/figure/agndiskflat/}}\resizebox{1.0\linewidth}{!}{\input{sims/figure/agndiskflat/densityradius_6_0.pgf}}
                \caption{Non-isothermal simulations with $\Sigma_0=\unit[30]{M_\odot/pc^2}$}
        \end{subfigure}\\
        \begin{subfigure}[b]{0.49\textwidth}
                
\renewcommand\pgfimage[2][NOSTD]{\includegraphics[#1.0]{sims/figure/agndiskisoflat/#sims/figure/agndiskisoflat/}}\resizebox{1.0\linewidth}{!}{\input{sims/figure/agndiskisoflat/densityradius_7_0.pgf}}
                \caption{Isothermal simulations with $\Sigma_0=\unit[50]{M_\odot/pc^2}$}
        \end{subfigure}
        \begin{subfigure}[b]{0.49\textwidth}
                
\renewcommand\pgfimage[2][NOSTD]{\includegraphics[#1.0]{sims/figure/agndiskflat/#sims/figure/agndiskflat/}}\resizebox{1.0\linewidth}{!}{\input{sims/figure/agndiskflat/densityradius_7_0.pgf}}
                \caption{Non-isothermal simulations with $\Sigma_0=\unit[50]{M_\odot/pc^2}$}
        \end{subfigure}
        \caption{Time evolution of the azimuthally averaged surface density profile of all the simulations performed.}
        \label{densityradius_large}
\end{figure*}
Without angular momentum transport, there can be no accretion. \refree{Therefore, such gas-poor regions are of great importance. Figure~\ref{densityradius_large} shows the time evolution of the azimuthally averaged density profiles of all isothermal and non-isothermal simulations. Gas-poor regions of different strengths can clearly be seen in all simulations.} The outer boundaries of these regions coincide roughly during almost the whole simulation with the position of an inner Lindblad resonance. Therefore, slightly disturbed gas loses its orbit and is pushed inward. High angular momentum material coming from smaller radii cannot cross this boundary to transport angular momentum outward \citep{2010MNRAS.407.1529H}. \\
The developments of the other simulations in figure~\ref{bhmass_0} show that most of them accrete  close to the Eddington limit for a much longer time.
Also, both simulations with $\Sigma_0=\unit[10]{M_\odot/pc^2}$ and weak cooling accrete more mass than the simulations with strong cooling.
This case of  hot accretion is discussed in the following section.
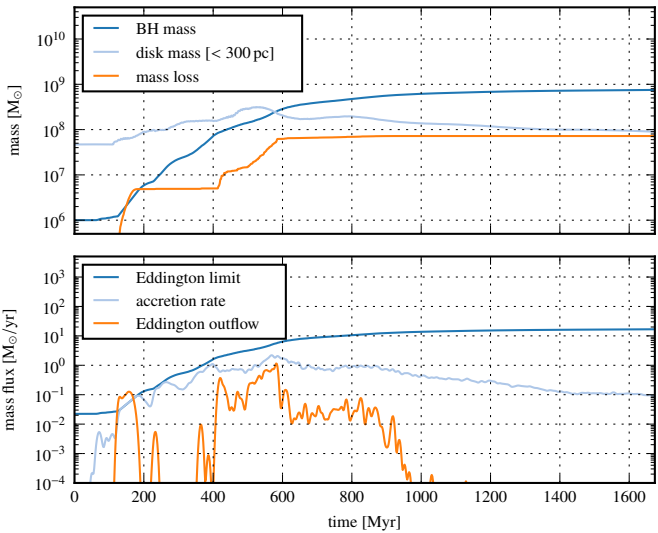
\figpgf{sims/figure/agndiskflat/3.E+1_4.5E-1_0.E-0_6.E+0/}{accdiskmass}{30_6_accdiskmass}{Time evolution of the accretion rate, Eddington outflow, and black hole mass of the simulation with $\Sigma_0=\unit[30]{M_\odot/pc^2}$ and $b_\mathrm{cool}=6$. Even before the bar potential is switched on completely ($\unit[186]{Myr}$), there is strong clumpy accretion. The inner disk mass rises quickly and  high accretion rates up to $\unit[2]{M_\odot/yr}$ are attained. Later on the Eddington outflow declines much more quickly than the accretion rate. This suggests a decrease in the clumpy accretion.}

\subsection{Hot accretion}
A typical case of the strong accretion and weak cooling is represented by the simulation with a surface density of $\Sigma_0=\unit[30]{M_\odot/pc^2}$ and the cooling parameter $t_\mathrm{cool}=6$.
Figure~\ref{30_6_accdiskmass} shows the accretion rate and the Eddington outflow as a function of time.
Even before the bar potential is completely switched on, substantial accretion onto the central object takes place.
Up to $\unit[600]{Myr}$ \LEt{ or: After 600 Myr ? }the accretion rate is close to the Eddington limit, while the outflow rate indicates an occasionally clumpy accretion.
The accretion rate reaches its maximum value of $\unit[2]{M_\odot/yr}$ after $\unit[600]{Myr}$. It then declines exponentially at a rate of $\unit[3]{/Gyr}$ until it settles down at a constant rate of $\unit[0.1]{M_\odot/yr}$ for the remaining $\unit[200]{Myr}$ of the simulation.
The course of the accretion rate seems to be closely correlated to the mass of the inner disk, for example  the maximum  inner disk mass of $\unit[3\cdot 10^8]{M_\odot}$ is reached shortly before the maximum accretion rate.
The inner disk passes the matter coming from outside to the
inner region with some delay.
The inner transport mechanism \replaced{conforms}{agrees} with  the previously described energy dissipation at strong shocks (sec. \ref{sec:tempstruct}), as illustrated in figure~\ref{streamspiral_0150}.
At the start of every simulation there are two shocks in the central region that dissipate kinetic energy and therefore generate a hot inner region with the gas spiraling in and approaching the center on eccentric orbits.
Thus, angular momentum can quickly be transported outward and mass can efficiently be accreted.\\
\refree{The hot accretion mode can also be distinguished from the cold accretion mode by the time evolution of the azimuthally averaged density profile (see  figure~\ref{densityradius_large}). The simulations that show the hot accretion mode have a high-density inner disk $<\unit[20]{pc}$ with a sharp outer boundary to larger radii. Additionally, the low-density region beyond $\unit[1]{kpc}$ is less pronounced as a result of the accretion flow from the kiloparsec scale.}

\subsection{Time evolution of black hole masses}
All simulations with cooling lead to the hot or cold accretion mode.
Which of the two is achieved does not depend on the initial mass, but instead on the cooling parameter.
Figure~\ref{bhmass_0} shows  an example of a mass-poor simulation with $\Sigma_0=\unit[30]{M_\odot/pc^2}$ where the exact cooling parameter is not important in the chosen parameter space, but only if it is below or above a specific critical value.
The final black hole mass depends on the initial disk mass and the accretion mode.
The resulting black hole masses of the cold and hot accretion modes differ in these simulations by around a factor of five.\LEt{ yes? }
The dependence of the final black hole mass on the initial disk mass does not change significantly between isothermal and non-isothermal simulations, as shown by the simulations with $b_\mathrm{cool}=3$ in figure~\ref{bhmass_0}.\\
The massive disk simulation with $\Sigma_0=\unit[50]{M_\odot/pc^2}$ and $b_\mathrm{cool}=3$, however, seems to be an exceptional case. It leads to an intermediate value for the final mass, about a factor of two more massive than in the simulation with $b_\mathrm{cool}=1$, but less massive than in the other simulations with the same initial disk mass. The accretion mode seems to be somewhere in between the cold and hot modes.
The threshold of the cooling parameter seems to be weakly dependent on the disk mass. In this case the transition region between the two modes has been exactly met.\\
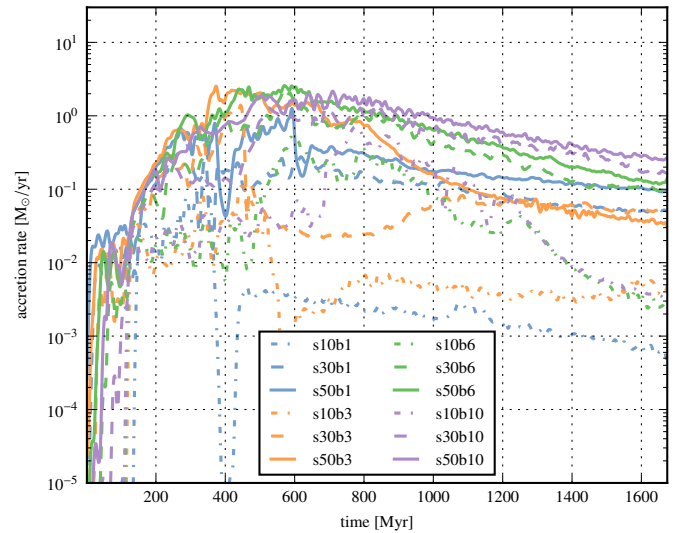
\figpgf{sims/figure/agndiskflat/}{accretionrate_0}{accretionrate_0}{Accretion rates of all galaxy simulations with cooling. The simulations with $\Sigma_0=\unit[10]{M_\odot/pc^2}$ initial surface density show considerably lower accretion rates than  the others at the end of the simulations. The massive simulations feature  exponential decreasing accretion rates of low variation from
$\unit[800]{Myr}$ onward.}
Figure~\ref{accretionrate_0} shows the accretion rates of all galaxy simulations with cooling. In contrast to the isothermal simulations there is no clear trend that  allows us to distinguish between strong and weak accretion accretion.
The simulations with $\Sigma_0=\unit[10]{M_\odot/pc^2}$ show an accretion rate of less than $\unit[5\cdot 10^{-3}]{M_\odot/yr}$ at the end of the simulation period.
Simulations, however, with higher initial surface densities exhibit peak accretion rates of more than $\unit[2\cdot 10^{-2}]{M_\odot/yr}$ before their accretion rates start to decline exponentially at about $\unit[800]{Myr}$.\\
Table~\ref{tab:agndiskflat} lists the mean and maximum accretion rates, as well as the final black hole masses.
Hot simulations with weak cooling and $\Sigma_0=\unit[30]{M_\odot/pc^2}$ reach higher black hole masses than the massive simulations with cold accretion.
All simulations with $b_\mathrm{cool}=10$ reach the highest black hole masses and accretion rates.
Table~\ref{tab:agndiskflat} also shows which structures in the inner disk form, and weather or not nuclear rings arise and persist for a long time.
This happens only for the least massive disks;   otherwise, gravitational instabilities set in which prohibit their formation at all or lead to their disintegration rather quickly.
In the case of strong cooling, the inner disk becomes gravitationally unstable and any large-scale structures that are generated initially in all simulations are destroyed immediately due to fragmentation.
Only in the hot simulations with weak cooling do these shocks persist throughout the whole simulation and do not become unstable in the inner regions.\\
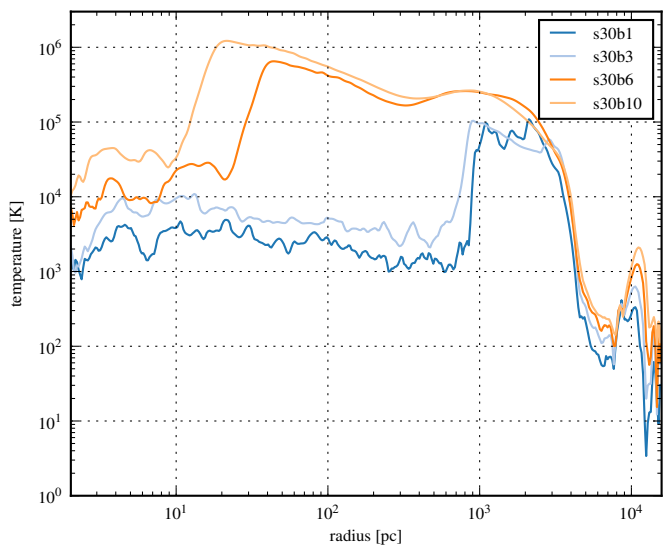
\figpgf{sims/figure/agndiskflat/}{temperature_6_0900}{temperature_6_0900}{Azimuthally averaged temperature curve of the non-isothermal simulations with $\Sigma_0=\unit[30]{M_\odot/pc^2}$. In the area $\unit[<1]{kpc}$, the simulations with strong cooling show temperatures that can typically be found in the warm ionized medium. In contrast, the simulations with weak cooling feature very high temperatures up to $\unit[10^6]{K}$.}
Figure~\ref{temperature_6_0900} reveals the radial azimuthally averaged temperature of the simulation with $\Sigma_0=\unit[30]{M_\odot/pc^2}$ for several cooling parameters.
The temperatures in the simulations with cooling match those with  cold accretion up to radii of $\unit[1]{kpc}$ to the warm ionized medium \citep{2015sfge.book....1G}.
This matches also within the area below about $\unit[10]{pc}$ in the case of simulations with hot accretion.
Beyond a $\unit[10]{pc}$ radius the simulations with hot accretion reach temperatures as high as $\unit[10^5-10^6]{K}$, which  corresponds to the typical temperatures of hot ionized gas \citep{ferriere_interstellar_2001}.
The temperatures in the isothermal simulations agree with molecular gas $\unit[\sim 100]{K}$  and the warm neutral or ionized medium \citep{ferriere_interstellar_2001}.

\subsection{Discussion}\label{sec:diskussionnoniso}
So far the involved temperatures in the simulations have not been discussed in detail because the cool function is independent of the temperature.
The temperature only goes  indirectly into the simulations by means of the gas pressure and the speed of sound, and thereby into the Toomre criterion, even though the speed of sound is a very important temperature dependent quantity that plays an important role in every hydrodynamic simulation.
All simulations with cooling initially encounter a strong shock that heats the gas inside  $\unit[1]{kpc}$ to very high temperatures of $\unit[10^{5\dots 6}]{K}$. \LEt{ is this standard? I think it should be expressed as  $\unit[10^5-10^6]{K}$; please check throughout  }
These temperatures can still be observed at the end in simulations with weak cooling.
Although very high temperatures up to $\unit[\sim 6\cdot 10^5]{K}$ can be obtained in the vicinity of the black hole \citep{2012NewAR..56...93A} it should be questioned whether this extends to the whole central region of a galaxy.
Additionally, it should be noted that hot ionized gas is generated by shock compression due to supernova explosions \citep{1977ApJ...218..148M}.
These events produce extended hot bubbles of low density in the interstellar medium \citep{2008MNRAS.389.1137S,2013MNRAS.429.1156S,2014A&A...566A..94K}. These might reach temperatures that are comparable to our simulation; however, they are spatially and temporally very confined. Therefore, their impact on the azimuthally averaged temperature will be small.\\
Overall, the processes that contribute to the cooling and heating of the gas have to be modeled in greater detail.
In this work only the contribution of shock and compression heating and a simple cooling in dependence on the dynamical timescale are considered.
The selection of the cooling parameter in our simulations is motivated by the the critical value $b_\mathrm{cool}$ \citep{2001ApJ...553..174G}, which is known to produce a gravitationally turbulent flow.
The actual cooling processes  run on timescales of $\sim10^{-6 \dots -4}\tau_\mathrm{dyn}$ \citep{2010MNRAS.407.1529H}.
Also, the heating by supernova explosions operates on timescales much shorter than the dynamical timescale (see section~\ref{sec:numsetupext}). 
Thus, the cooling employed in this work should model the effective contribution of the balance of all actual cooling and heating processes.
\replaced{The leveraged hydrodynamics software \textsf{FOSITE} \citep{2009CoPhC.180.2283I}}{\textsf{FOSITE}} \LEt{ The FOSITE instrument?, The FOSITE computer program? if this is an acronym it should be spelled out here }provides the opportunity of cooling using the method from \cite{1990ApJ...351..632H}, which models a gray radiation with mean Rosseland opacities according to \cite{1994ApJ...427..987B}.
The key issue is that the complete gas is immediately cooled down to the minimum allowed temperature  (see section ~\ref{sec:numsetupext}). 
This is inevitable because no explicit heating has been included, \replaced{since}{even though} a model of the heating processes is extremely difficult.
The interstellar medium is predominantly heated by bright young stellar clusters and supernovae \citep{2000MNRAS.318..227K}.
Supernovae can be heated extremely well by means of shock compression and momentum input and are especially important for the inner galaxy disk areas of high density, which exhibit high opacities and star formation (e.g., \citealp{2013ApJ...769..100S,2015ApJ...802...99K}), although these are processes that cannot be resolved in the simulations carried out in the present work.
Therefore, modeling the heating processes appropriately requires a subgrid-scale model that accounts for star formation (radiation input) and supernova feedback.
The result is gas turbulence on smaller scales than the mesh of the hydrodynamic simulations.
In order to model these movements on the smallest scales,  turbulent pressure can be introduced \citep{2010ApJ...709..191M,2012NewAR..56...93A}, which together with the thermal pressure results in  effective pressure \citep{2005ApJ...622L...9S,2005MNRAS.361..776S,2006ApJ...645..986R}.
This effective pressure is sufficient to prevent  collapse on small scales, and increases the Toomre parameter and the Jeans length.
For example,  \cite{2010ApJ...709..191M} use  turbulent sound velocities on the order of $\unit[10-100]{km/s}$, which are consistent with observations (e.g., \citealp{1998ApJ...507..615D}).
The high sound velocities found in the simulations in this work can therefore be interpreted as effective sound velocities, which are composed of thermal and turbulent pressure and which motivated  \cite{2012ApJ...747...60K} to use high isothermal sound velocities of up to $\unit[20]{km/s}$.\LEt{ yes? }

If we consider the high temperatures obtained in the simulations to be realistic values,  this would imply the existence of ionized gas. Hence, magnetic fields could play an important role in  hydrodynamics.
\cite{2012ApJ...751..124K} performed two-dimensional non-self-gravitating magnetohydrodynamic \LEt{ yes? used only here so needs to be spelled out } simulations of barred galaxies. They report that all simulations show nuclear rings and that the typical spiral arms have lower densities than in hydrodynamic simulations.
Also, initially there is a second bar \citep{1989Natur.338...45S} inside  the nuclear ring in accordance with observations (e.g., \citealp{1995MNRAS.274..369S,1996A&AS..118..461F}).
Furthermore, \replaced{\cite{2012ApJ...751..124K}}{Shlosman et al. (1989)} \LEt{ yes? } report mean accretion rates of about $\unit[10^{-4\ldots -2}]{M_\odot/yr}$.
Since they do not account for self-gravity of the gas disk nor do they consider the influence of an outer spiral structure on the potential, they do not obtain a persistent mass flow from the outer regions onto the central disk after a short initial accretion phase.

\section{Impact on AGN evolution}
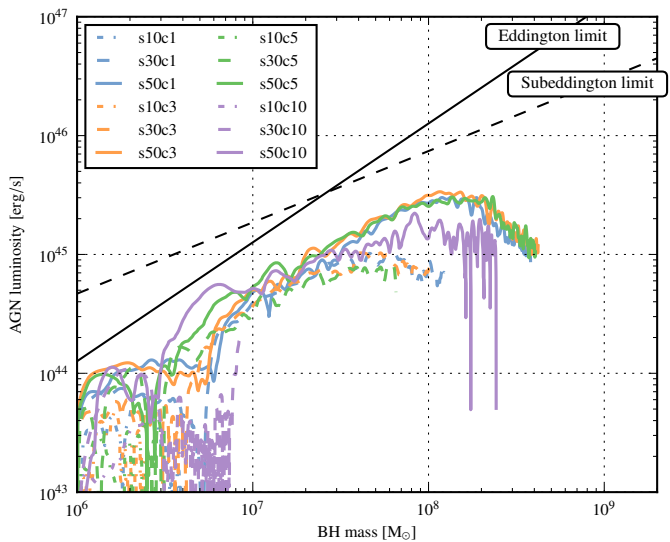
\figpgf{sims/figure/agndiskisoflat/}{evolutionpath_0}{isoevolutionpath_0}{Evolutionary tracks of the black holes in the isothermal simulations. Since only the massive simulations lead to appreciable black hole masses, the mass-poor simulations remain in the lower left corner. The other simulations move on a path parallel to the Eddington limit (black line), then they bend and feature a constant accretion rate. The dashed black line indicates the sub-Eddington limit for redshifts of $0.6 < z< 0.8$ from \cite{2010MNRAS.402.2637S}. It does not pose a strong limit for these simulations.}
The gathered black hole masses and accretion rates can be combined to evolutionary tracks of the different black holes in an AGN luminosity-mass-diagram (cf. \cite{shen_biases_2008,2010MNRAS.402.2637S}). If we assume a black hole accretion efficiency \citep{2011ApJ...728...98D} of $\eta=0.1$, the AGN luminosity can be calculated as a function of the accretion rate $\dot{M}$
\begin{equation}
L_\mathrm{acc} = \eta \dot{M} c^2.
\end{equation}
Figure~\ref{isoevolutionpath_0} shows that only simulations with $\Sigma_0=\unit[30]{M_\odot/pc^2}$ or $\Sigma_0=\unit[50]{M_\odot/pc^2}$ result in noteworthy evolutionary tracks.
The mass-poor  simulations nearly vanish in this diagram because only very low black hole masses could be reached.
The other simulations proceed partially along the Eddington boundary, but half a decade below  because of the above-mentioned effect of clumpy accretion and because the  inner boundary of the computational domain is \replaced{at a comparatively large radius}{comparatively large}.\LEt{ I am sure I\ have changed your meaning here, I apologize   }
A disproportionately high amount of mass is lost by means of Eddington outflow.
All the massive simulations eventually deviate from the track along the Eddington boundary.
In the beginning they accrete more slowly and later decrease further.
Unfortunately, the simulation time is not long enough to follow until the accretion is finished  and the accretion rate collapses completely.\\
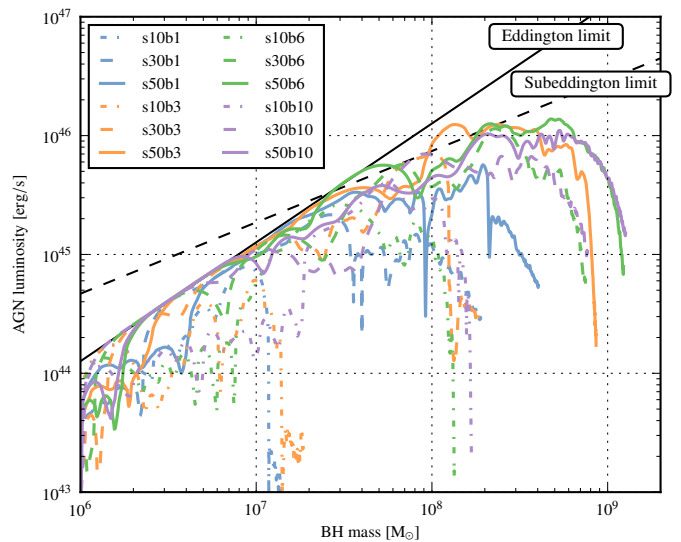
\figpgf{sims/figure/agndiskflat/}{evolutionpath_0}{evolutionpath_0}{Evolutionary tracks for the black holes of galaxy simulations with cooling. All simulations accrete for a long time near the Eddington limit (black line). Depending on the initial conditions they leave this evolutionary path at different black hole masses and then show strongly declining AGN luminosities. The dashed black line indicates the sub-Eddington limit for redshifts of $0.6 < z< 0.8$ from \cite{2010MNRAS.402.2637S}.}
In contrast to the isothermal case the simulations with cooling show considerable black hole growth. Thus, figure~\ref{evolutionpath_0} of the AGN luminosity--mass plot shows long evolutionary tracks for all simulations with cooling.
The evolutionary tracks all look  very similar: First, an accretion phase near the Eddington limit takes place, which can be strongly fluctuating.
This could be caused by the remaining distance of the black hole to the inner boundary of the computational domain, as has been explained in detail in section \ref{sec:accandagnev}.
After some time, depending on the parameters, the simulations stop accreting at the Eddington limit and bend down to gradually weaker AGN luminosities.
Often the decline is very abrupt so that the black hole only grows marginally.

\cite{2010MNRAS.402.2637S} plot the luminosities of more than $60000$ quasars against their black hole masses depending on their redshift.
They observe that quasars with low black hole masses of up to $\unit[10^{7}]{M_\odot}$ can reach luminosities as high as their respective Eddington luminosities, but higher mass quasars are subject to a second boundary below the Eddington luminosity (cf. \citealp{2006ApJ...648..128K,2015ApJ...811..148C,2014ApJ...796...25S}).
The observed ensemble is bounded from below by the detection limit at $\unit[10^{45}]{erg/s}$.
Unfortunately, this limits the comparability to the isothermal simulations. Here only the most massive simulations reach luminosities of more than $\unit[10^{45}]{erg/s}$ since the maximum AGN luminosity, which is achieved for an accretion rate at the Eddington limit, is proportional to the black hole mass. Therefore, additional simulations reaching higher black hole masses would be desirable. 
In agreement with
the observations  the  simulations are below the sub-Eddington limit, but without a larger sample of simulations we cannot give statistically relevant evidence.\\
The youngest quasars, which reach  black hole masses comparable to those  in our non-isothermal simulations (up to $\unit[2\cdot 10^{9}]{M_\odot)}$, have a redshift of $z\approx 0.8$.
We plot in figure~\ref{evolutionpath_0} the sub-Eddington limit from \cite{2010MNRAS.402.2637S} for quasars with a redshift of $0.6<z<0.8$ (black dashed line).
The results of the simulations comply with this limit very well before their accretion rates quickly decline.
The argument would become stronger if simulations with higher initial disk masses were available.

\section{Conclusions}
In this work we have analyzed isothermal and non-isothermal multi-scale simulations of the gas flow in barred spiral galaxies with particular attention to the growth of the central black hole. In the case of isothermal disks we could show that the simulations evolve in two completely distinct ways depending on the speed of sound and the initial mass of the gas disk.
In the strong accretion mode the inner accretion disk ($r<\unit[300]{pc}$) gains more than $\unit[10^8]{M_\odot}$, which causes the inner disk to become gravitationally unstable, thus leading to very efficient accretion.\\
In the weak accretion mode, however, the mass $M$ of the inner disk is too low to drive a graviturbulent accretion. Instead, kinetic energy is dissipated at large-scale shock fronts, which leads to gas flow on eccentric and inward spiraling orbits, driving a weak accretion flow.
In agreement with observations, nuclear rings and spirals can be observed \added{at least} for transitory periods in the simulations. Often these regular structures are destroyed because the inner disk gains mass and becomes gravitationally unstable.\\
In the second part of this paper we extended the models by accounting for energy transport, shock heating, and radiative cooling. Again, we were able to identify two modes of accretion in the simulations that are, however, completely different from  the isothermal case. In the cold accretion mode the accretion rate is generally lower. Since a sufficient cooling is given, a marginally stable accretion disk is formed,  but the accretion is mostly depleted after a short time because no further material is supplied from the kiloparsec region.\\
For the hot accretion mode, two remarkable shocks reach from the inner rim of the simulation domain up to the kiloparsec region. At these shocks energy is dissipated and the gas spirals inward on highly eccentric orbits. This is a very efficient accretion mechanism with typical accretion rates of $\unit[10^{-1\dots 0}]{M_\odot}$. The assigned AGN evolutions are in very good agreement to the sub-Eddington limit, but a physical explanation for this phenomenon remains hidden and requires the exploration of a larger parameter space.\\
Of course our models show some opportunities for improvement. The fixed gravitational potential of the stellar component might change over the  course of such long-running simulations. The energy budget in the non-isothermal simulations is based on a very simple cooling model, which lacks the inclusion of various feedback mechanisms. Also, if the disk becomes hot, the vertical pressure scale-height will increase. In this case the approximation as a geometrical thin disk has to be dropped. \\
In comparison to \cite{2012ApJ...747...60K} our results show the importance of self-gravity for the dynamical evolution of the gas disk. While they did not report overall significant black hole accretion, we were able to follow the evolution of the black hole for much longer (nine bar rotations) and found sustained high accretion rates. In many simulations graviturbulent accretion modes or filamentary structures were observed in the inner accretion disk. For this reason it is important to resolve the central accretion disk until the gravitational potential of the black hole dominates and the orbits become round with Keplerian velocities.  \cite{2012ApJ...747...60K} have also measured gas inflow for several simulations, but they note that simulations with black hole masses below $\unit[10^8]{M_\odot}$ suffer from non-circular orbits near the inner boundary of the computational domain, which gives rise to high gas inflow rates. This emphasizes that resolving at least some part of the Keplerian accretion disk around the central black hole, even for its initial mass, is important.
While we have not investigated the evolution of the nuclear ring in detail, our results suggests that it becomes gravitationally unstable for the more massive gas disks. As a consequence star formation in these regions seems to be likely \citep{2014IAUS..303...43K}. Thus, including star formation models and feedback mechanisms would be worth further investigation.\\
The comparison of the calculated AGN evolutions to the observed Eddington and sub-Eddington limit shows good agreement with the simulations. The non-isothermal simulations indeed show considerable AGN evolution accreting at the Eddington limit, and a decline at higher black hole masses where the sub-Eddington limit is observed. The isothermal simulations are also consistent with the observations, but often reach black hole masses that are below the detection limit for their corresponding AGN luminosities.

\bibliography{paper}

\begin{appendix}
                
        \section{Potential of a triaxial homeoid}\label{sec:barpotential}
        \cite{1984A&A...134..373P} characterized the potential of a triaxial homeoid following the original publication by \cite{ferrers1887} and \cite{1969efe..book.....C}.
\LEt{ see note 12. There are only two references here, so "and" and no comma are necessary }        Since this is rather complex, we note the result for the special case $n=1$ used in this work so that it can easily be reproduced. 
        Let the density distribution of the considered body be
        \begin{equation}
        \rho = \begin{cases}\rho_\text{bar}\left(1-g^2\right)^n &\mbox{if } g < 1,\\ 0 &\mbox{else.}\end{cases}
        \end{equation}
        with $g=\sfrac{x^2}{a^2}+\sfrac{y^2}{b^2}+\sfrac{z^2}{c^2}$\replaced{: $a$ the major and $b$, $c$ the minor semi-axes;}{  $a$, $b$, and $c$ the semi major and minor axes;} \LEt{ see note 4 (and make sure the format is the same in both cases)  }  the potential in the $z=0$ plain is given by
                \begin{align}
        \Phi_\text{bar}\left(x,y,0\right) = &-\frac{\pi G a b c \rho_\text{bar}}{2}\nonumber\\
        &\cdot\left(\right. W_{000}\nonumber\\
        &\quad+x^2\left(x^2 W_{200} +2 y^2 W_{110}-2 W_{100}\right)\nonumber\\
        &\quad\left.+y^2\left(y^2 W_{020}-2 W_{010}\right)\right)
        \end{align}
        Using the incomplete elliptic integral of the first and second kind \citep{1965hmfw.book.....A},
        \begin{align}
        F\left(\varphi,k\right) &= \int\nolimits_0^\varphi \frac{1}{\sqrt{1-k^2\sin^2\theta}}\dd \theta\\
        E\left(\varphi,k\right) &= \int\nolimits_0^\varphi\sqrt{1-k^2\sin^2\theta}\dd \theta,
        \end{align}
        we define the following parameters:
        \begin{align}
        k &= \sqrt{\frac{a^2-b^2}{a^2-c^2}}\\
        \varphi &= \arcsin\left(\sqrt{\frac{a^2-c^2}{a^2+\lambda}}\right)\\
        W_{000} &= \frac{2}{a^2-c^2} F\left(\varphi,k\right)\\
        W_{100} &= \frac{2}{\left(a^2-b^2\right)\cdot\sqrt{a^2-c^2}}\left(F\left(\varphi,k\right)-E\left(\varphi,k\right)\right)\\
        W_{001} &= \frac{2}{a^2-b^2}\sqrt{\frac{b^2+\lambda}{\left(a^2+\lambda\right)\left(c^2+\lambda\right)}}\nonumber\\
        &\quad-\frac{2}{\left(b^2-c^2\right)\sqrt{a^2-c^2}} E\left(\varphi,k\right)\\
        W_{010} &= \frac{2}{\Delta\left(\lambda\right)} -W_{100}-W_{001}\\
        W_{110} &= \frac{W_{010}-W_{100}}{a^2-b^2}\\
        W_{011} &= \frac{W_{001}-W_{010}}{b^2-c^2}\\
        W_{101} &= \frac{W_{100}-W_{001}}{c^2-a^2}\\
        W_{200} &= \frac{1}{3}\left(\frac{2}{\Delta\left(\lambda\right)\cdot\left(a^2+\lambda\right)}-W_{110}-W_{101}\right)\\
        W_{020} &= \frac{1}{3}\left(\frac{2}{\Delta\left(\lambda\right)\cdot\left(b^2+\lambda\right)}-W_{011}-W_{110}\right).
        \end{align}
        Here $\lambda$ has to satisfy the equation
        \begin{equation}
        \frac{x^2}{a^2+\lambda}+\frac{y^2}{b^2+\lambda} = 1.
        \end{equation}
        The solution is calculated by  means of root finding and depends on the coordinates $x$, $y$.
\end{appendix}
\end{document}